\documentclass[traditabstract]{aa}  
\usepackage{graphicx}
\usepackage{natbib}
\usepackage{amsmath}             
\usepackage{txfonts}

\newcommand{\ratioo} {N({\rm H}_2) / I_{\rm CO}}

\newcommand{\kms}   {{\rm \  km \  s^{-1}}}
\newcommand{\K}   {{\rm \  K}}

\newcommand{\Htwo} {\rm H_{2}}
\newcommand{\Xunit} {\,{\rm cm^{-2}/(K\kms)}}
\newcommand{\cmt} {\,{\rm cm^{-2}}}
\newcommand{\cc} {\,{\rm cm^{-3}}}

\newcommand{\HI}{\ion{H}{i}}

\newcommand{\msun}{M$_{\odot}$}   
   
\def\hcop{HCO$^+$}
\def\hh{H$_{\rm 2}$}
\def\DJ{$\Delta${\it J}}
\def\DF{$\Delta${\it F}}
\def\tkin{$T_{\rm kin}$}
\def\scm{cm$^{-2}$}
\def\sqpc{pc$^{-2}$}
\def\ccm{cm$^{-3}$}
\def\pow#1#2{#1$\times$10$^{#2}$}

\begin{document}

\title{ Dense gas and star formation in the Outer Milky Way }
		
\author{J. Braine\inst{1} \and Yan Sun\inst{2}  \and Y. Shimajiri\inst{3} \and F.F.S. van der Tak\inst{4,5} \and  Min Fang\inst{2} \and Ph. Andr\'e\inst{6} 
\and Hao Chen\inst{7}\and Yu Gao\inst{8,2} }

\institute{Laboratoire d'Astrophysique de Bordeaux, Univ. Bordeaux, CNRS, B18N, all\'ee Geoffroy Saint-Hilaire, 33615 Pessac, France \\
             \email{jonathan.braine@u-bordeaux.fr}
        \and Purple Mountain Observatory, Chinese Academy of Sciences, 10 Yuanhua Road, Nanjing 210023, China     \and
            Kyushu Kyoritsu University, Jiyugaoka 1-8, Yahatanishi-ku,Kitakyushu, Fukuoka, 807-8585, Japan \and
            SRON Netherlands Institute for Space Research, Landleven 12, 9747 AD Groningen, The Netherlands \and
            Kapteyn Astronomical Institute, University of Groningen, NL \and
            Laboratoire d'Astrophysique (AIM), CEA/DRF, CNRS, Universit\'e Paris-Saclay, Universit\'e Paris Diderot, Sorbonne Paris-Cit\'e, 91191 Gif-sur-Yvette, France \and
            Research Center for Intelligent Computing Platforms, Zhejiang Laboratory, Hangzhou 311100, China \and
            Department of Astronomy, Xiamen University, Xiamen, Fujian 361005, China
        }
       
\date{Received xxxx; accepted xxxx}

\abstract { We present maps and spectra of the HCN(1--0) and HCO$^+$(1--0) lines in the extreme outer 
Galaxy, at galactocentric radii between 14 and 22 kpc, 
with the 13.7 meter Delingha telescope.  The 9 molecular clouds were selected from a CO/$^{13}$CO survey of the outer quadrants.
The goal is to better understand the structure of molecular clouds in these poorly studied subsolar metallicity regions and the relation with star formation.
The lines are all narrow, less than $2 \kms$ at half power, enabling detection of the HCN hyperfine structure in the stronger sources and allowing us to 
observationally test hyperfine collision rates.  The hyperfine line ratios show that the HCN emission is optically thin with column densities estimated at 
$N$(HCN) $ \approx 3 \times 10^{12}$\scm.
The HCO$^+$ emission is approximately twice as strong as the HCN (taken as the sum of all components), in contrast with the inner Galaxy and nearby galaxies where they are similarly strong. 
For an abundance ratio $\chi_{HCN}/\chi_{HCO^+} = 3$, this requires a relatively low density solution for the dense gas, with n(\hh) $\sim 10^3 - 10^4$\ccm.  The $^{12}$CO/$^{13}$CO line ratios are similar to solar neighborhood values, roughly 7.5, despite the low $^{13}$CO abundance expected at such large radii.  The HCO$^+$/CO and HCO$^+$/$^{13}$CO integrated intensity ratios are also standard at about 1/35 and 1/5 respectively.  The HCN is weak compared to the CO emission, with HCN/CO $\sim 1/70$ even after summing all hyperfine components.  In low-metallicity galaxies, the HCN deficit is attributed to a low [N/O] abundance ratio but in the outer disk clouds it may also be due to a low volume density.  
At the parsec scales observed here, the correlation between star formation, as traced by 24~$\mu$m emission as is standard in extragalactic work, and dense gas via the HCN or HCO$^+$ emission, is poor, perhaps due to the lack of dynamic range.
We find that the lowest dense gas fractions are in the sources at high galactic latitude
($b>2^\circ$, $h > 300$pc above the plane), possibly due to a lower pressure.
  }	


\maketitle

\section{Introduction}

A general theory of star formation needs to be tested in multiple environments.
Most work \citep[see e.g.][]{Kennicutt12} has studied either 
galactic disk star formation, 
star formation in bright and perturbed environments such as mergers or ULIRGs, or bright low metallicity
objects such as blue dwarf galaxies.
Observations of the outer Galaxy provide a new environment for the study of the star formation 
cycle (\HI\ to $\Htwo$ to stars).
The outer disk is cooler and has a lower gas content and in particular less molecular gas than the brighter 
regions mentioned above.
The star formation rate is low \citep[see e.g. ][]{Sodroski97} and the metallicity is subsolar 
but not extremely low in the outer disk \citep[gradient of $\approx 0.04$ dex kpc$^{-1}$,][]{Pedicelli2009}.
The morphology and gravitational potential are those of a rotating disk, unlike mergers and many dwarfs.
Finally, the outer disk, defined as beyond the R$_{25}$ radius, is typically where the gas surface density exceeds 
that of stars (see e.g. Figure 7 of \citet{Kennicutt12} or Figure 1 of \citet{Hoekstra01}).  For a population 
of sun-like stars, a brightness of 25 magnitudes per square arcsecond, defining the R$_{25}$ radius, corresponds 
to a stellar surface density of 6.6\msun \sqpc.
Hence, the outer Galaxy represents a new environment (where $\Sigma_{\rm gas} \ga \Sigma_*$) for the study of the star formation cycle.

A further motivation comes from the observations and simulations showing that galaxies continue to be 
fueled by inflowing gas \citep[see][ and references therein]{Linsky2003,Schmidt2016}, particularly when a bar 
(i.e. a non-axisymmetric potential) is present as is the case for the Galaxy.
Thus the outer disk represents a part of the future of our Galaxy. 

The star formation rate (hereafter SFR) at kpc scales is well-known to be linked to the presence of gas and 
particularly molecular gas \citep[e.g.][]{Kennicutt89}.  Closer to cloud scales, the cloud life-cycle (pre-SF, embedded 
SF, exposed SF, cloud dispersal) is such that much more scatter is present in the SFR-$\Htwo$ relation.
More recently, the link between the dense gas mass, generally as traced by the high-dipole moment molecules
HCN or HCO$^+$, has been shown to be approximately linear 
in galactic disks at both kpc \citep[e.g.][]{Gao04,Gao04b,Kepley2014,Chen2017,Braine17}
and pc scales \citep[e.g.][]{Wu2010,Shimajiri2017} with approximately the same SFR/M$_{gas}$ ratio \citep[see Fig. 13 of ][]{Jimenez2019}.
In galactic centers, the SFR/M$_{gas}$ ratio is clearly lower than in disks \citep[see lower panel of Fig. 7 in][]{Chen2015, Usero2015}. 
Table 5 and Figure 13 of \citet{Jimenez2019} show that over about 10 orders of magnitude in $L_{\rm HCN}$, there is a scatter of 
0.37 dex in the $L_{\rm IR}$/$L_{\rm HCN}$ ratio, where $L_{\rm IR}$ traces the SFR, in agreement with \citet{Gallagher2018}.
 
IRAS Far-IR observations and the follow-up in CO by \cite{Wouterloot89} revealed molecular gas out to a 
galactocentric distance of about 18~kpc.  
Perhaps surprisingly, \citet{Digel94} found several molecular 
clouds beyond 18~kpc but they are all much less CO-bright (although of comparable mass) than Orion
viewed at a similar distance and resolution.
The atomic gas extends considerably further \citep[e.g.][]{Kalberla2009} but with little or no evidence of star formation 
and young stars.

Recent increases in mapping speed and receiver bandwidth combined with post-IRAS Far-IR 
surveys such as Herschel Hi-Gal \citep{Molinari2016} and the APEX Atlasgal \citep{Schuller09} has led to 
revived interest in the outer Galaxy.  \citet{Sun15} surveyed a vast region 
covering $100 < l < 150^\circ$ and $-3 < b < 5.25^\circ$ in the $J=1-0$ transition of CO and isotopologues to determine 
how much  molecular gas is present at large galactocentric distances.  In addition to previously known clouds, 
they found 49 previously undetected extreme outer Galaxy clouds ($R_{gal}>14$kpc). The CO emission is generally quite 
weak, with integrated intensities below $10 \K \kms$ at the CO peaks. 
$^{13}$CO(1--0) and C$^{18}$O(1--0) were observed simultaneously and the images and spectra 
are presented in \citet{Sunyanthesis}.  $^{13}$CO(1--0) emission is detected in nearly 
half of the sources.  The $^{12}$CO/$^{13}$CO line ratio is typically about 5 in peak temperature 
and 7 in integrated intensity; the C$^{18}$O lines were not detected (see Sect. 4.1). 

 \citet{Sunyanthesis} started a comparison between star formation and the presence of molecular gas in the outer Galaxy.
 One of the results of her thesis was that while star formation was not always associated with 
 molecular gas in these extreme outer Galaxy clouds, the clouds hosting young stellar objects (YSOs) 
 were generally more massive than those without YSOs.
Here we pursue the comparison between tracers of star formation and molecular gas in the outer Galaxy, 
focussing on the link between {\it dense} gas and star formation.  At kpc scales, there is a good correlation between
the dense gas mass and the star formation rate \citep{Chen2015} as for whole galaxies \citep{Gao04}.
The low-$J$ transitions of the high dipole moment molecules HCN and HCO$^+$ 
are standard tracers of dense gas because high densities (close to $10^5$ \ccm) are required for collisional excitation \citep{Evans99}.  
However, the lines are much (factor 10--100) weaker than the CO lines so
the existing $^{12}$CO(1--0) and $^{13}$CO(1--0) observations were used to define regions to be observed
in HCN and HCO$^+$ in the extreme outer Galaxy.  The dataset obtained is used to explore the physical conditions of 
the molecular medium far out in the Galaxy and the link between dense gas and star formation in a new environment \citep[e.g.][]{Glover2012}. 
There is evidence that HCN is more sensitive to metallicity variations than HCO$^+$ \citep{Rudolph2006,Nishimura2016b,Braine17}.
The outer Galaxy has a subsolar metallicity but also a generally weak UV field and thus provides 
complementary information to observations of low metallicity galaxies, which often have
high UV-fields.

Only a few observations of dense gas tracers in the outer Galaxy have been made \citep{Yuan2016} 
so we significantly increase this sample.  \citet{Yuan2016} observed a large sample of Planck Galactic Cold Clumps, 
detected in the sub-mm range by the Planck satellite and generally close to the Sun.  The samples are 
quite different in that the \citet{Sun15} survey was blind and {\it none} of their 72 sources  were 
detected in the C$^{18}$O line.  Very recently, \citet{Fontani2022} observed molecular lines, including 
HCN and \hcop, in a sample of actively star-forming
IRAS-selected cores and \citet{Patra2022} mapped the HCN and \hcop(1-0) lines around regions with massive stars,
also in the outer Galaxy.  These two samples are quite different from ours, as we will see (Sect. 4.2).

The paper is structured as follows: after presenting the observations and the basic results, we estimate the excitation 
temperatures, optical depths, and column densities of the CO lines and then for the HCN, where the satellite line 
strengths are used to estimate the optical depth.  In this part, we initially follow \citet{Yuan2016}.  
The following section presents radiative transfer calculations of 
the HCN line ratios and also the HCN/HCO$^+$ line ratio, in order to constrain the volume density and determine 
whether the CO-emitting gas can also be responsible for the HCN and \hcop\ line emission.  We then discuss, 
with in particular the goal of understanding why the line ratios we measure are not typical of galactic or most 
extragalactic observations, and summarize the results for this rather unique data set. 

\begin{figure}
	\centering
	\includegraphics[width=\hsize{}]{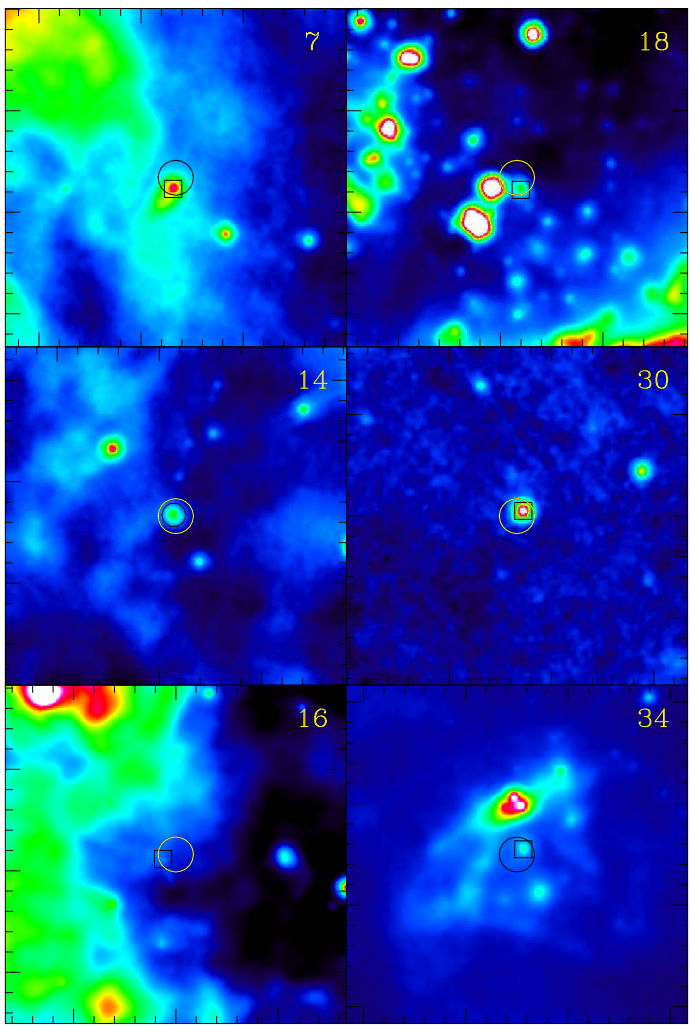}
	\caption{ WISE 22 $\mu$m band images of sources 7, 14, 16, 18, 30, and 34 (identification in upper right corner).  These are the sources for which only a single pointing is available.  The images are all 10 arcminutes in size, oriented in $l$ and $b$.  The central circle gives the HCN/HCO$^+$ beamsize.  The transfer function (rainbow from black to white) of the WISE observations differs from image to image. For cloud 7 it is from 91.5 to 98; for cloud 14 it is 100.5 to 107; for 16 it is 103 to 115; for 18 it is 102 to 114; for 30 it is from 108 to 112 and for cloud 34 it is from 108 to 135 counts. Following  \citet{Sunyanthesis}, class I sources within our maps are shown as circles and class II as squares.}
	\label{images} 
\end{figure}

\begin{figure}
	\centering
	\includegraphics[width=\hsize{}]{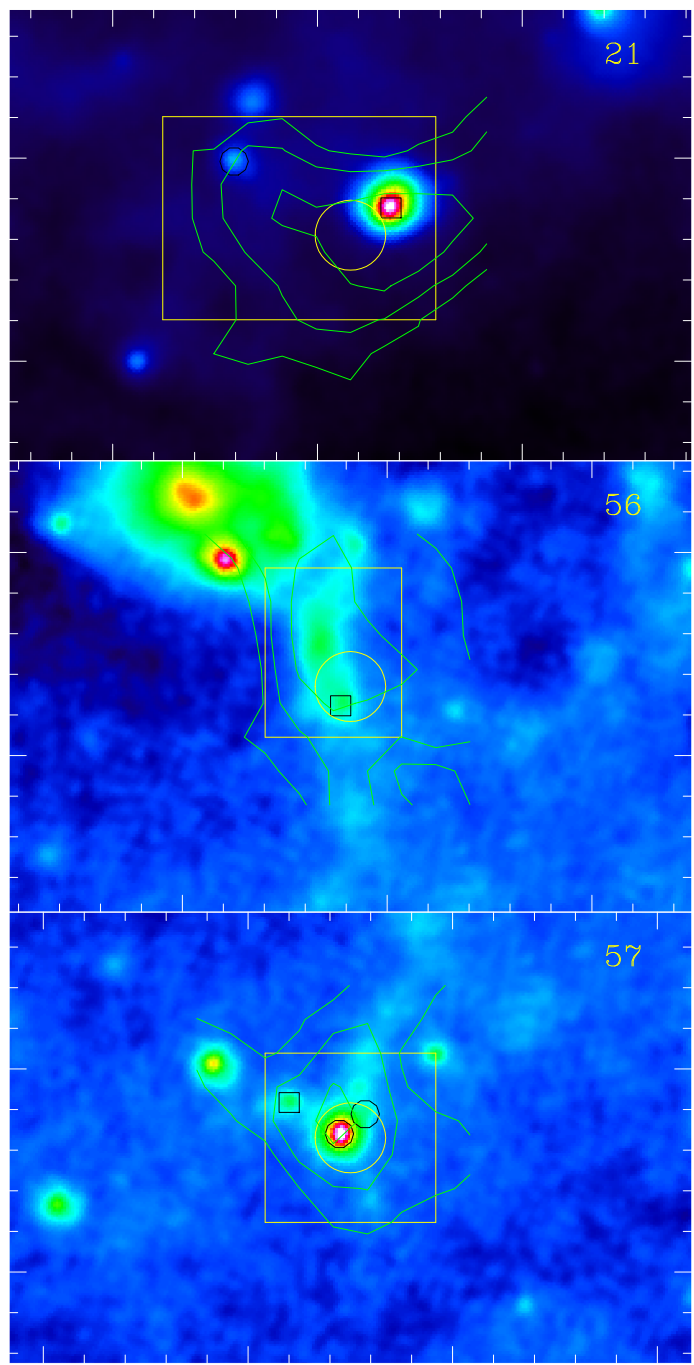}
	\caption{ WISE 22 $\mu$m band images of sources 21, 56, and 57 (identification in upper right corner).  These are the sources which were  mapped.  The images are all $600" \times 400"$ in size, oriented in $l$ and $b$.   The color scale (rainbow from black to white) of the WISE observations differs from image to image. For cloud 21 it is from 95 to 116 and for clouds 56 and 57 the range is from 129 to 134 counts.   The central circle gives the HCN/HCO$^+$ beamsize and is the (0,0) position in Figures \ref{map21}, \ref{map56}, and \ref{map57} which show the spectra.The box indicates the size of the region shown in Figures \ref{map21}, \ref{map56}, and \ref{map57}. Following  \citet{Sunyanthesis}, class I sources within our maps are shown as circles and class II as squares.}
	\label{im215657} 
\end{figure}

\begin{figure}
	\centering
	\includegraphics[width=\hsize{}]{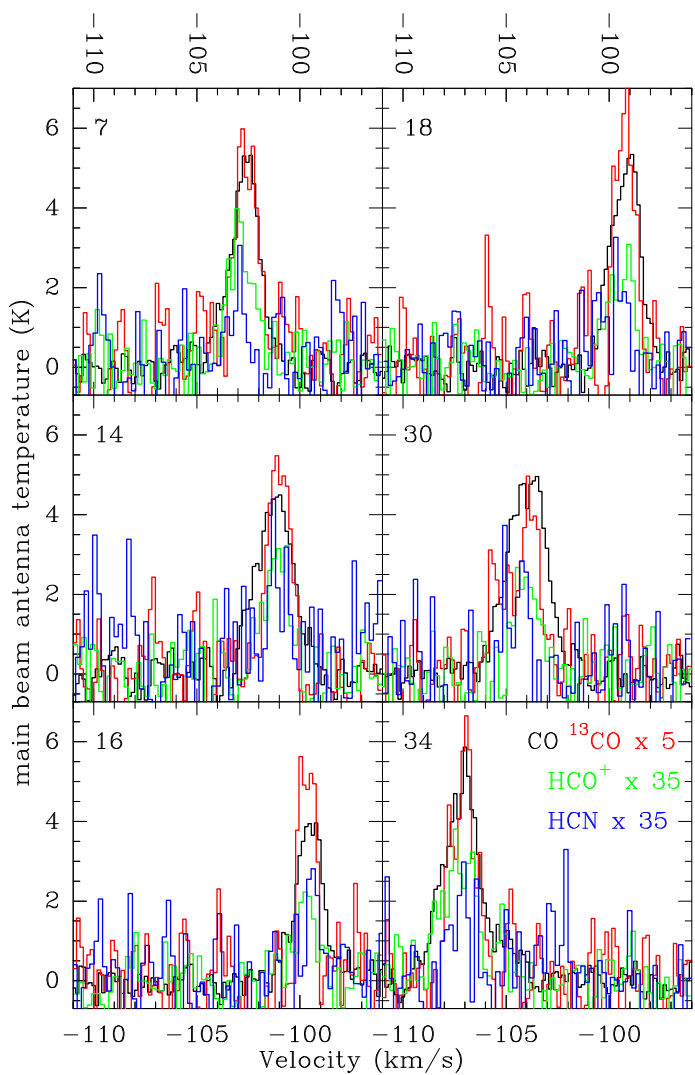}
	\caption{Spectra for the clouds that have not been mapped, as in Fig.~\ref{images}.  The axes are the same for each cloud.  
	$^{12}$CO is in black and corresponds to the temperature scale.   $^{13}$CO is in red and has been multiplied by 5.  HCO$^+$, in green, has been multiplied by 35.  The HCN spectra have been multiplied by 35 for comparison with HCO$^+$ but also multiplied by 9/5 to account for the satellite lines as explained in the text.  The $^{12}$CO and $^{13}$CO spectra have been convolved with a gaussian kernel to the angular resolution of the HCN and HCO$^+$ spectra.}
	\label{indiv} 
\end{figure}

\begin{figure*}
	\centering
	\includegraphics[width=\hsize{}]{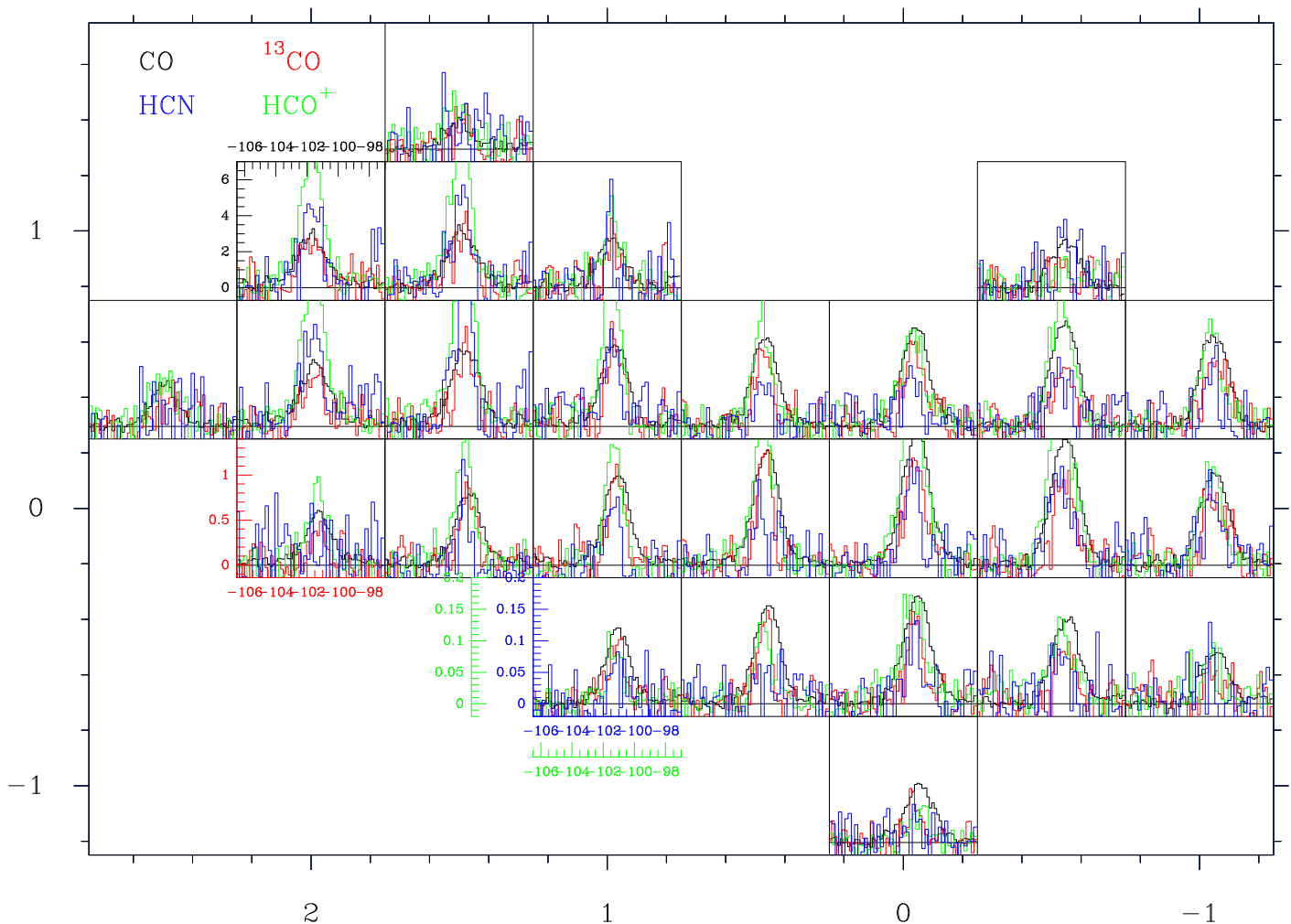}
	\caption{Spectra of Cloud 21.  Positions are indicated on the $x$ and $y$ axes as offsets in arcminutes and each box represents 30$"$.  $^{12}$CO is in black, $^{13}$CO in red, HCO$^+$ in green and HCN in blue.  The velocity scale goes from $-106.5\kms$ to $-97\kms$.  The temperatures scales are -.7 to 7K for $^{12}$CO, -.14 to 1.4K for $^{13}$CO, -.02 to .2 for HCO$^+$ and for HCN.  
	The HCN line (only central component shown) has been multiplied by 9/5 so that the line intensity represents the total line flux for comparison with other species.  The $^{12}$CO and $^{13}$CO spectra have been convolved to the same angular resolution as the HCN and HCO$^+$.}
	\label{map21} 
\end{figure*}

\begin{figure*}
	\centering
	\includegraphics[width=\hsize{}]{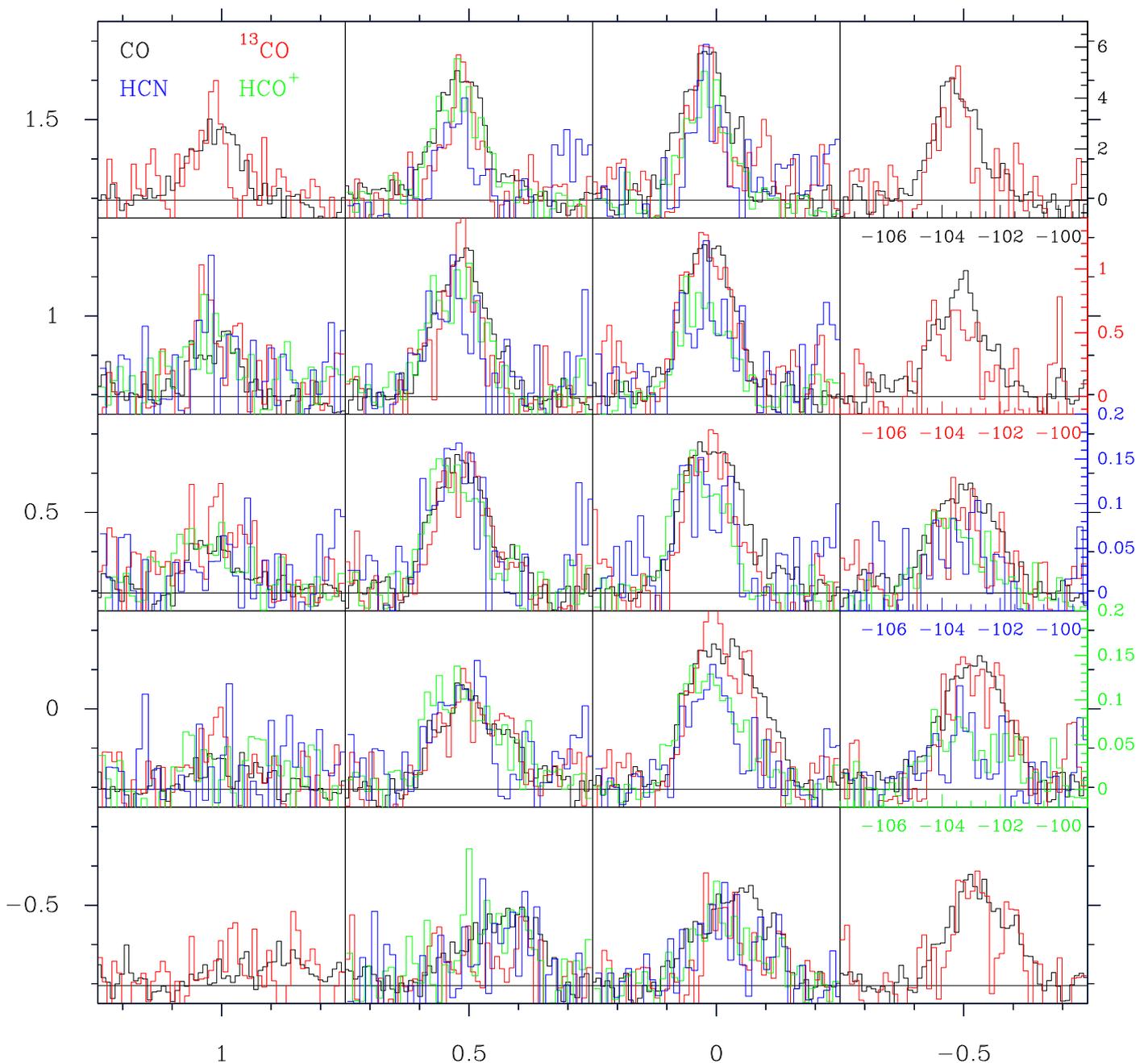}
	\caption{Spectra of Cloud 56.  Positions are indicated on the $x$ and $y$ axes as offsets in arcminutes and each box represents 30$"$.  $^{12}$CO is in black, $^{13}$CO in red, HCO$^+$ in green and HCN in blue.  The velocity scale goes from $-107.5\kms$ to $-99\kms$.  The temperatures scales are -.7 to 7K for $^{12}$CO, -.14 to 1.4K for $^{13}$CO, -.02 to .2 for HCO$^+$ and for HCN.  The HCN line (only central component shown) has been multiplied by 9/5 so that the line intensity represents the total line flux for comparison with other species. The spectra in the corners show only CO as not all positions were observed in HCN and HCO$^+$.}
	\label{map56} 
\end{figure*}

\begin{figure*}
	\centering
	\includegraphics[width=\hsize{}]{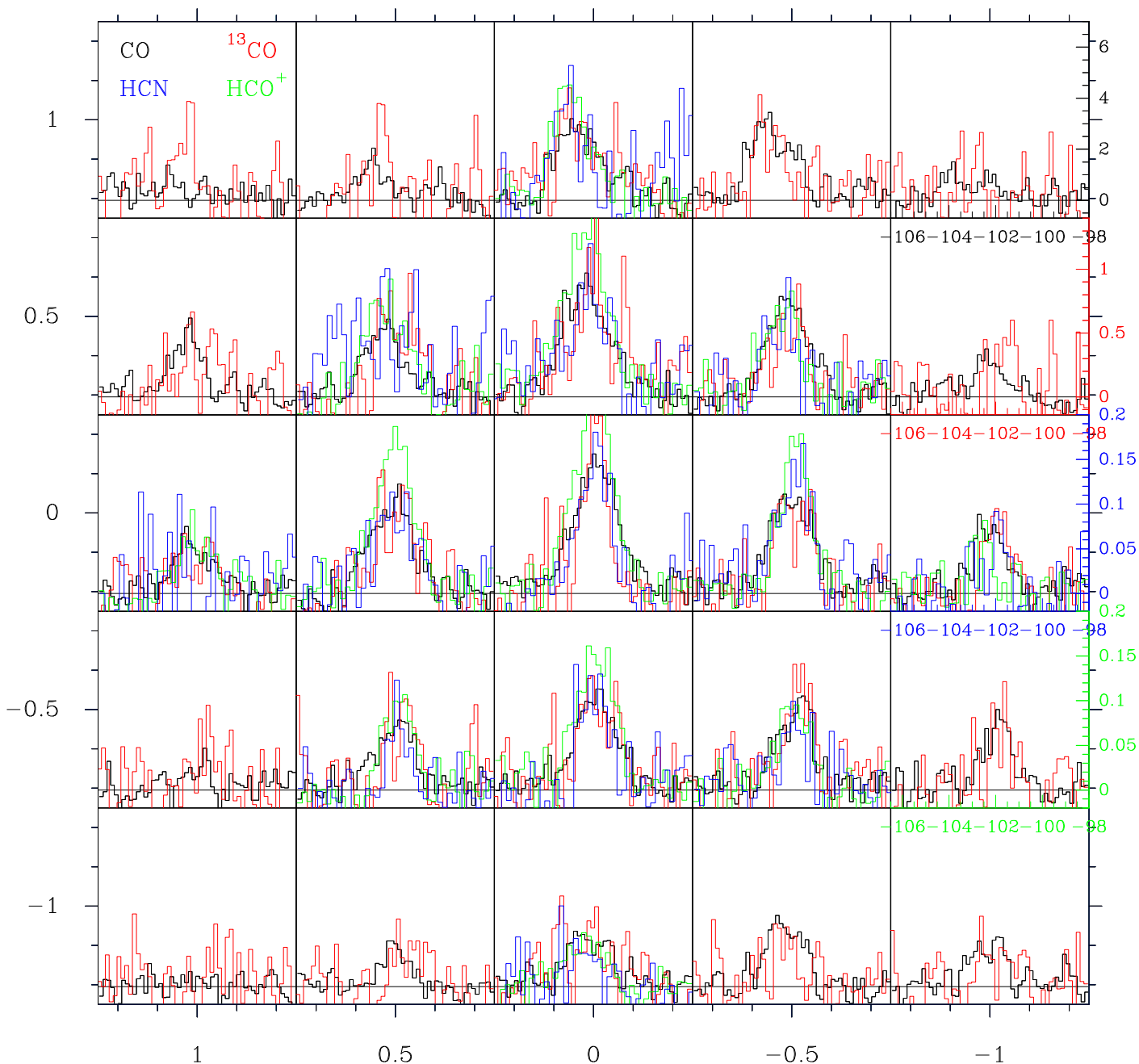}
	\caption{Spectra of Cloud 57.  Positions are indicated on the $x$ and $y$ axes as offsets in arcminutes and each box represents 30$"$.  $^{12}$CO is in black, $^{13}$CO in red, HCO$^+$ in green and HCN in blue.  The velocity scale goes from $-106.5\kms$ to $-98\kms$.  The temperatures scales are -.7 to 7K for $^{12}$CO, -.14 to 1.4K for $^{13}$CO, -.02 to .2 for HCO$^+$ and for HCN.  The HCN line (only central component shown) has been multiplied by 9/5 so that the line intensity represents the total line flux for comparison with other species. The spectra in the corners show only CO as not all positions were observed in HCN and HCO$^+$.}
	\label{map57} 
\end{figure*}

\begin{figure}
	\centering
	\includegraphics[width=\hsize{}]{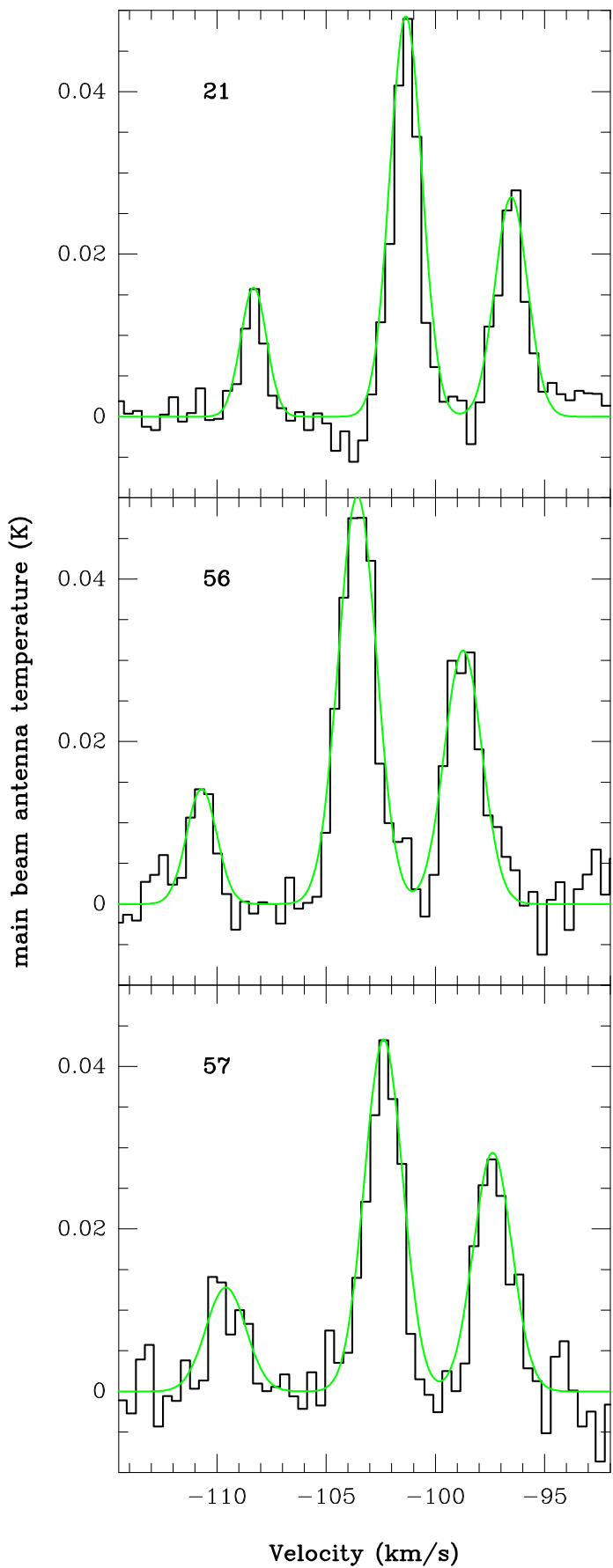}
	\caption{ Sum of spectra for each of the clouds for which we were able to make maps.   
	The lines are narrow although each spectrum represents a sum of many positions. The clear separation of the hyperfine components enables us to estimate the optical depth. The parameters of the gaussian fits shown in green are given in Table 2.}
	\label{hyper} 
\end{figure}

\begin{figure}
	\centering
	\includegraphics[width=\hsize{}]{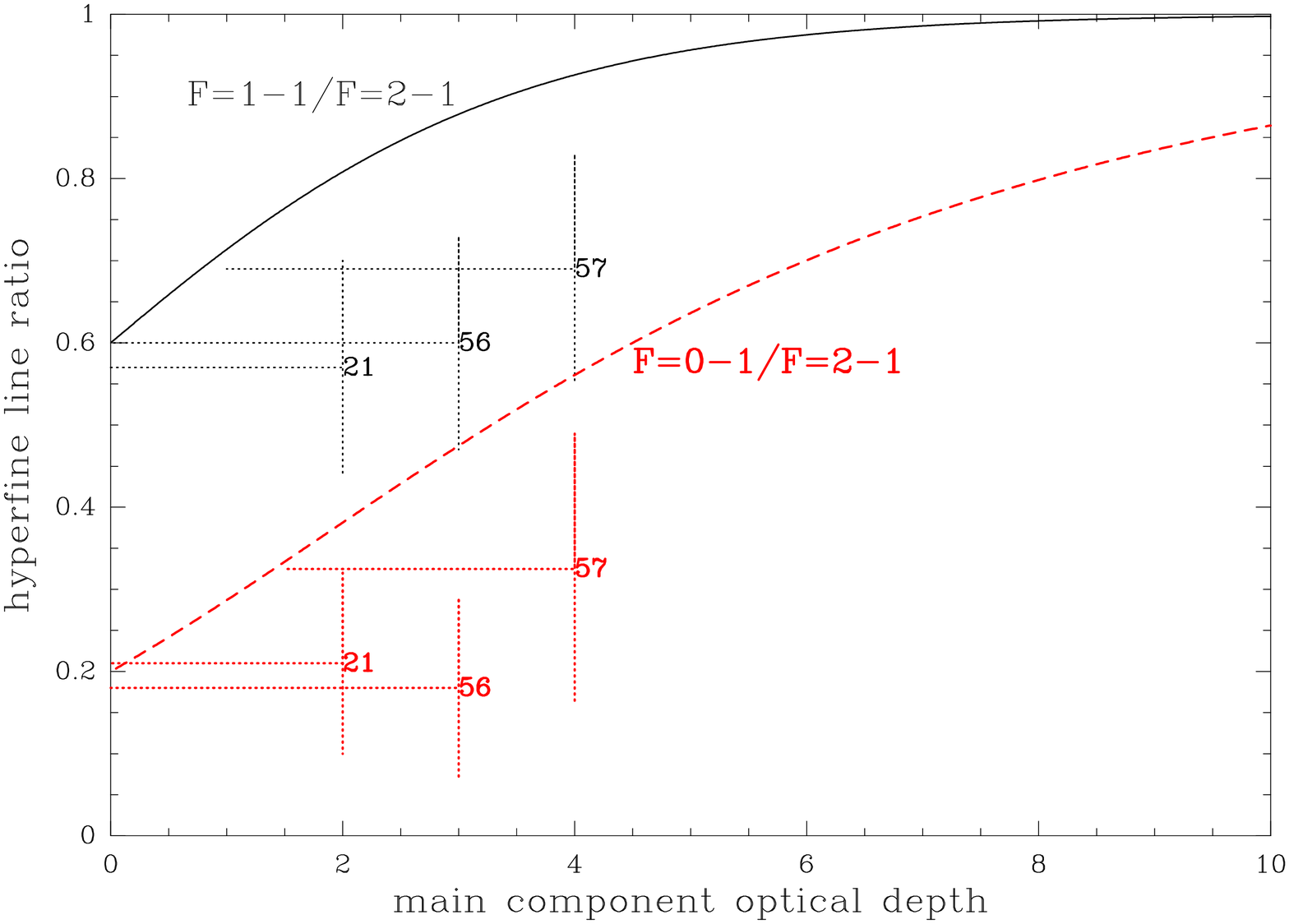}
	\caption{ Local Thermodynamic Equilibrium representation of hyperfine component ratios as a function of main component (F=2-1) optical depth. 
	The observed ratios for clouds 21, 56, and 57 are indicated. The vertical dashed lines indicate the uncertainties from Table 3.}
	\label{naive} 
\end{figure}

\begin{figure}
	\centering
	\includegraphics[width=\hsize{}]{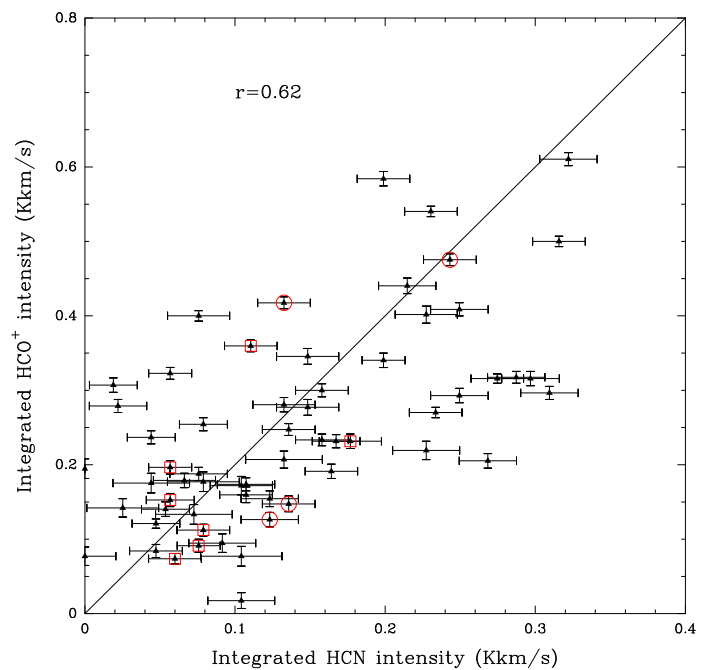}
	\caption{ Integrated intensities of HCN and HCO$^+$ lines for all positions observed.  All data are on the main beam temperature scale and the central HCN intensity has been multiplied by 9/5 to account for the satellite hyperfine components.  The line indicates $I_{\rm HCO^+} = 2 I_{\rm HCN}$, close to the best-fit value of 1.92.  Class I and II sources are indicated by respectively red circles and squares.}
	\label{hcn_area} 
\end{figure}

\begin{figure}
	\centering
	\includegraphics[width=\hsize{}]{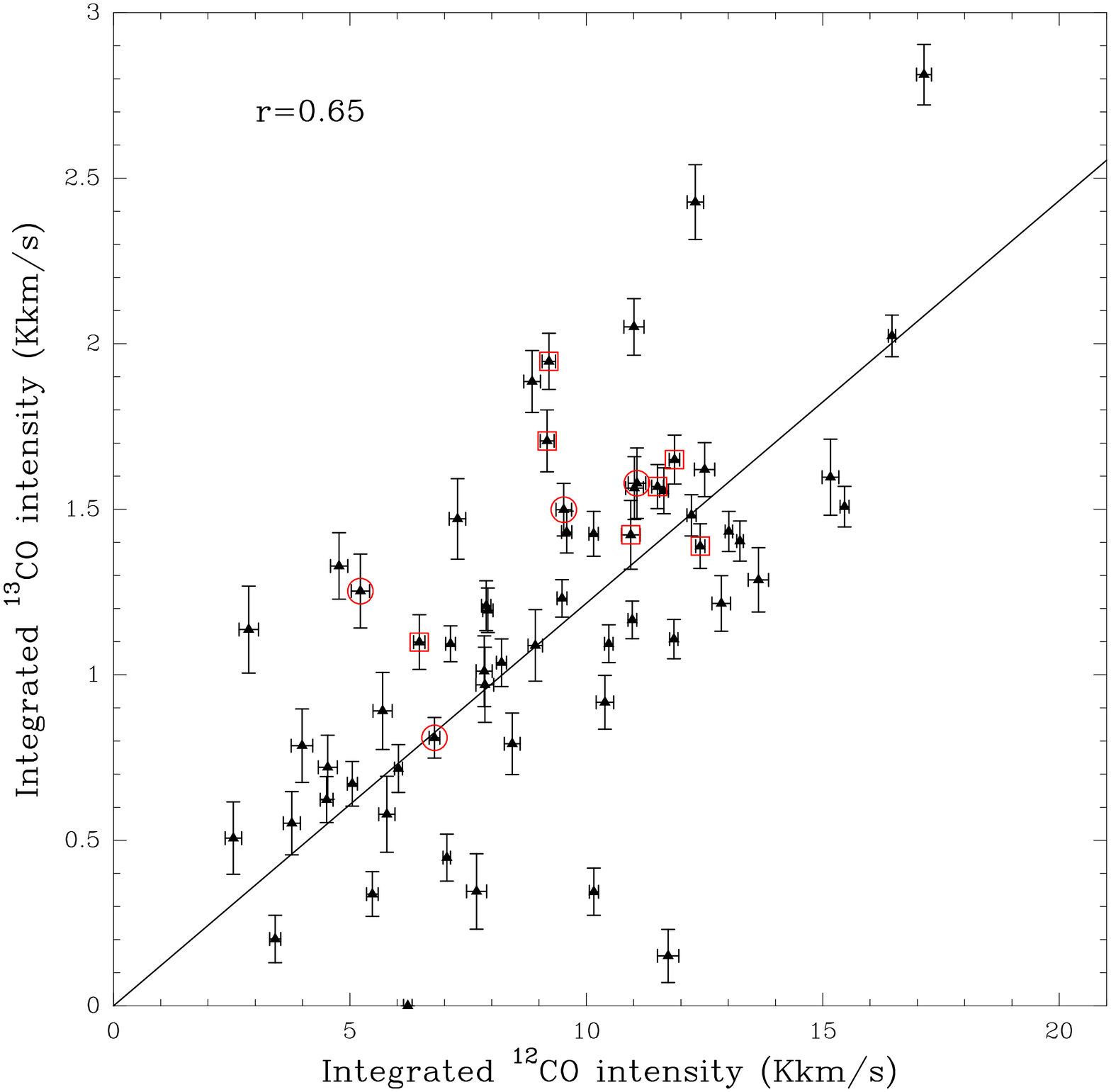}
	\caption{ Integrated intensities of $^{12}$CO and $^{13}$CO lines for all positions observed.    
	The line indicates the best-fit: $I_{^{13}{\rm CO}} = \frac{I_{\rm CO}}{8}$.  Class I and II sources are indicated by respectively red circles and squares.}
	\label{co_area} 
\end{figure}

\section{Cloud sample and Observations}

The sources were selected from the \citet{Sun15} sample of outer Galaxy CO clouds.  All of the clouds are quite far from the galactic center, ranging from 14 to 22 kpc with the \citet{Reid14} rotation curve.  Although a different rotation curve will yield slightly different results, these are clearly far outer Galaxy objects, close to or beyond twice the solar circle distance.  Sources were chosen to have ($a$) an integrated CO(1--0) intensity $I_{\rm CO} >7$ K km s$^{-1}$ and ($b$) $I_{^{13}\rm CO} \ge1.4$ K km s$^{-1}$ and ($c$) a peak $^{13}$CO temperature above 0.7 Kelvin.  The $^{13}$CO data are from \citet{Sunyanthesis} and the cloud characteristics, $^{12}$CO, and $^{13}$CO intensities are provided in Table 1.  This yielded a sample of nine clouds: 7, 14, 16, 18, 21, 30, 34, 56 and 57, all of which have $^{13}$CO sizes between 2 and 5 square arcminutes and peak CO temperatures above 4 Kelvins. 
The central positions of all clouds were observed and clouds 21, 56, and 57, the brightest in HCN and HCO$^+$, were mapped.
Fig.~\ref{images} presents WISE 22 $\mu$m band images, supposed to trace star formation, with the positions observed and symbols indicating class I and class II YSOs.  Fig.~\ref{im215657} shows the same for the three clouds that were mapped in HCN and \hcop\ along with contours of CO intensity.  The WISE 22 $\mu$m band\footnote{https://irsa.ipac.caltech.edu/Missions/wise.html} was chosen as it was the only tracer of star formation available for all sources. 

The YSOs within the clouds were identified with the criteria described in \citet{Koenig2012} using the infrared data from the Two Micron All Sky Survey \citep[2MASS:][] {Skrutskie2006} and the Wide-field Infrared Survey Explorer \citep[WISE:][]{Wright2010}. We restrain our sources to those with photometric errors less than 0.1 mag for 2MASS data and signal-to-noise ratio greater than 5 for WISE data. The criteria in \citet{Koenig2012} are designed to search for young stars at Class~I and Class~II stages employing two methods. The first is based on photometry in the WISE 3.4, 4.6 and 12~$\mu m$ bands. The contamination from extragalactic sources (star-forming galaxies and AGNs), shock emission blobs, resolved PAH emission objects can be removed according to their locations in the [3.4]$-$[4.6] vs. [4.6]$-$[12] color-color diagram and their WISE photometry. For the sources not detected in the WISE 12~$\mu m$ band, YSOs are identified from the dereddened $K_{\rm s}-$[3.4] vs. [3.4]$-$[4.6] color-color diagram. In our work, the extinction used to deredden the photometry is estimated from its location in the $J-H$ vs. $H-K_{\rm s}$ color-color diagram as described in \citet{Fang2013}.  

The nominal central positions observed correspond to the CO maxima and not necessarily a peak in star formation, as can be seen in Figures \ref{images} and \ref{im215657}.
Below we give the assessment of the star formation for each observed region \citep{Sunyanthesis}, starting with the single positions and followed by the maps.\\
{\it Cloud 7}: An IRAS point source is present as well as a class II source within the DLH beam.\\
{\it Cloud 14}: No IRAS point source within the beam but  a class I source at beam center.\\
{\it Cloud 16}: No IRAS point source but there is a class II source within the DLH beam.\\
{\it Cloud 18}: No IRAS point source but there is a class II source within the DLH beam.\\
{\it Cloud 30}: No IRAS point source but there is a class II source within the DLH beam.\\
{\it Cloud 34}: Class II source within DLH beam and IRAS source nearby (see Fig. \ref{images}).\\
{\it Cloud 21}: IRAS point source and class II source at edge of central beam but within map, which also covers a class I source.  Map shows no obvious difference in  line ratios at positions with star formation.\\
{\it Cloud 56}: No IRAS point source within but a class II source near nominal center.  Star formation (22 $\mu$m emission) extends to higher $b$ and HCN/HCO$^+$ emission appears to follow.\\
{\it Cloud 57}: IRAS point source at center and two class I sources, one coincident with the IRAS source. A class II source is also present further from the center.\\
Class I sources are indicated as circles in Figures \ref{images} and \ref{im215657}.  Class II are shown as squares, following  \citet{Sunyanthesis}.

The observations were carried out in May 2017 with the 13.7m telescope of the Purple Mountain Observatory
(PMO) in Delingha, Qinghai, China, hereafter DLH.  The beamsize at the HCN and HCO$^+$ $J=1-0$ frequencies is 1$'$ which corresponds to a linear scale of 3~pc at the typical distance of 10~kpc of our clouds (except clouds 56 and 57 which are further such that the beam size is about 4.3~pc).  
The Superconducting Spectroscopic Array Receiver system, a nine-beam (3X3beam, with separation of 3 arcmin), sideband- separating receiver (Shan et al. 2012) was used as front end. The fast
fourier transform spectrometer with a bandwidth of 1 GHz provides 16384 channels and
a spectral resolution of 61 kHz \citep[see details in ][]{Shan2012}, leading to a velocity resolution of 0.2 km s$^{-1}$.
The telescope is at an altitude of 3200~meters and the weather was good during the observations, leading to system temperatures of 140 -- 150K.  

The standard chopper-wheel calibration was adopted to obtain the antenna temperature, $T_{\rm A}^{*}$ \citep{Kutner81}. 
Spectral intensities are further converted to a scale of main-beam brightness temperature using 
$T_{\rm MB} =T_{\rm A}^{*}/\eta_{MB}$, where the main-beam efficiency 
$\eta_{\rm MB}$ is 0.57 at $\sim$90 GHz (see the status report of the 13.7m telescope~\footnote{http://english.dlh.pmo.cas.cn/fs/}). 
Considering the small angular size of our targets traced by CO, and the smaller region traced by dense gas, the position-switching mode was used. 
Two beams with a separation of 6 arcmin were used to observe on-source pointing and emission-free OFF position simultaneously. The pointing accuracy is estimated to be better than 5 arcsec. The mapping step is 30 arcsec.

\section{Results}

\subsection{Maps and Spectra}

Table 1 provides source numbers, positions, $^{12}$CO intensities, galactocentric distance R$_{\rm gal}$, heliocentric distance, all from \citet{Sun15},  and  $^{13}$CO intensities from \citet{Sunyanthesis}.  The CO data are on the main beam scale (i.e. corrected for telescope efficiency) and come from observations with the Delingha telescope which has a beam size of 48$"$ at the CO(1--0) frequency.

In Galactic observations, several clouds can be found along the line of sight at different velocities.  More common in the inner Galaxy, this is also the case for these outer Galaxy observations.  The outer Galaxy CO survey \citep{Sun15,Sunyanthesis} covered not only the peaks but a vast region in the outer Galaxy and thus maps of each cloud.
The spectra and maps of the clouds in the appendix of \citet{Sunyanthesis} show that generally the CO and $^{13}$CO is strongest in the distant cloud we are interested in and that even when this is not the case (only cloud 30), the clouds can be visually separated in the maps.  

\begin{table}
\begin{center}
\begin{tabular}{lllllll}
Cloud \# & $l$ & $b$ & I$_{\rm CO}$ & R$_{\rm gal}$ & dist &  I$_{\rm ^{13}CO}$ \\
 & $\circ$ & $\circ$ & K km s$^{-1}$ & (kpc) & (kpc)  & K km s$^{-1}$   \\
\hline
7 & 104.983 & 3.317 & 11.4 & 14.5 & 9.9 & 1.6      \\
14 & 109.292 & 2.083 & 11.5 & 14.6 & 9.5 & 1.7      \\
16 & 109.500 & 2.608 & 7.4 & 14.5 & 9.3 &  1.4     \\
18 & 109.792 & 2.717 & 10.3 & 14.4 & 9.2  & 1.8      \\
21 & 114.342 & 0.781 & 18.0 & 15.0 & 9.5   &  2.4     \\
30 & 121.817 & 3.050 & 13.7 & 16.4 & 10.3 & 2.0      \\
34 & 122.775 & 2.525 & 12.8 & 17.1 & 11.0   &  1.9     \\
56 & 137.759 & -0.983 & 17.6 & 21.7 & 14.7  &  3.1     \\
57 & 137.775 & -1.067 & 12.2 & 21.4 & 14.4  &  1.6     \\
\end{tabular}
\caption[]{Cloud sample: source numbers, positions, $^{12}$CO intensities, galactocentric distance R$_{\rm gal}$, heliocentric distance, all from \citet{Sun15},  and  $^{13}$CO intensities from \citet{Sunyanthesis}. }
\end{center}
\end{table}

The new observations of the HCN(1--0) and HCO$^+$(1--0) lines are shown along with the pre-existing $^{12}$CO(1--0) and  $^{13}$CO(1--0) lines in Figures \ref{indiv},  \ref{map21}, \ref{map56}, and \ref{map57}.  The temperature scales are the same from one figure to another so that the figures can be straightforwardly compared.  
We immediately see that the HCN(1--0) and HCO$^+$(1--0) lines were detected in all sources and that the \hcop\ line is systematically stronger than the HCN line.  
Central velocities are very similar (generally $\pm$ one channel) and the optically thick $^{12}$CO line is generally broader (by $\sim40$\%) than the dense gas tracers \citep[compare Table 2 with Table 1 of ][]{Sun15}. 

Table 2 presents the results of the HCN and \hcop observations.  The integrated intensity, velocity, and line width are given for the 
central spectra in the \hcop(1-0) line and for the main line of HCN(1-0).  Clouds 21, 56, and 57 were partially mapped in HCN and \hcop and the lines indicated by {\it sum} are the fits to the co-added (i.e. summed)  spectra.  The summed spectra have considerably higher signal-to-noise ratios (see last column) so the individual hyperfine components were fit and the results are presented in the same way. 

\subsection{HCN hyperfine structure}

HCN and \hcop are considered to be tracers of dense gas due to their high electric dipole moment.
Their critical densities, where collisional de-excitation equals spontaneous de-excitation \citep[see Eq. 4 and Table 1
of][]{Shirley15}, are $n_{crit,HCN(1-0)} \approx 4 \times 10^5$ \ccm\ and  $n_{crit,HCO^+(1-0)} \approx 6 \times 
10^4$ \ccm.  These densities are roughly 2 orders of magnitude higher than for CO so the typical molecular cloud 
densities at which CO is collisionally excited do not suffice to generate detectable HCN or \hcop\ emission.
The outer galaxy environment is cool, with a weak radiation field, and as we shall see the HCN lines are 
optically thin or nearly so.

Due to the nuclear quadrupole moment of Nitrogen ($^{14}$N), there is hyperfine structure which is observable 
in the ground-state rotational transition ($J=1 \rightarrow 0$).  Three lines are present and the line ratios 
can be used to estimate optical depth.  The main line is the $F=2 \rightarrow 1$ transition at 88.63185 GHz, 
with satellite lines $F=1 \rightarrow 1$ and $F=0 \rightarrow 1$ shifted by respectively 4.8 and 
$-7 \kms$.  Figure \ref{hyper} shows the co-added HCN(1-0) 
spectra for clouds 21, 56, and 57.  The individual spectra do not have a sufficiently high signal-to-noise ratio to 
measure the HFS at individual positions, particularly for the weakest component.   The spectra are presented 
for each cloud separately in Figures \ref{map21}, \ref{map56}, and \ref{map57}, which 
show that the co-added spectra are from regions with strong CO  (and $^{13}$CO) emission. The second-strongest component 
is clearly visible in the individual spectra of cloud 56 (Fig. \ref{map56}) which has the lowest noise level (see Table 2).  
Generally, when spectra are stacked, the velocity is recentered because the stacking concerns gas at different 
velocities -- this shifting introduces additional uncetainties because of the way the velocity difference is determined. 
This is not the case here which is why we say ``summed`` or ``co-added`` rather than stacked.
 
 \subsection{Relative line intensities}

As can be seen in Figures \ref{hcn_area} and \ref{co_area}, the CO and $^{13}$CO, and HCN and \hcop\ line fluxes are correlated.
The relation is approximately  $I_{\rm HCO^+} \approx 0.03 I_{\rm CO} \approx 0.21 I_{^{13}{\rm CO}} \approx 2 I_{\rm HCN}$ for the set of data.  The correlation coefficients are $r=0.62$ between dense gas tracers and $r=0.65$ between the $^{12}$CO and $^{13}$CO lines.  Mixing the dense and total gas tracers,
for respectively CO-HCO$^+$, CO-HCN, $^{13}$CO-HCO$^+$, and $^{13}$CO-HCN, the correlation coefficients are 0.53, 0.44, 0.44, and 0.43.
Hence, the total gas mass (clouds) and dense gas mass (cores) are linked but the dense gas tracers (HCN, \hcop ) are more strongly correlated between each other, as are the general molecular gas tracers (CO, $^{13}$CO), than between clouds and cores.
 
\section {Physical conditions from an LTE analysis}

In this section we apply the standard formulae to estimate optical depths and column densities of the species presented here.
The formulae are presented in \citet{Yuan2016}, \citet{Mangum2015}, and \citet{Shirley15} and the works cited therein.
We initially follow the reasoning in \citet{Yuan2016} for straightforward comparison and then present another approach.  
Without making ad hoc hypotheses, it is difficult to make non-LTE (local thermodynamic equilibrium) calculations so
the following section presents radiative transfer calculations with RADEX \citep{vandertak07}.  

The observations are of clouds between 14 and 22 kpc from the Galactic center so we expect abundances 
to be subsolar \citep[e.g.][]{Pedicelli2009} and we follow the \citet{Pineda2013} 
$^{12/13}$C gradient \citep[see Eq. 4 in][]{Yuan2016} and the \citet{Fontani2012} $^{12}$CO abundance gradient 
 \citep[see Eq. 7 in][]{Yuan2016}.

\subsection{Molecular column densities: $^{12}$CO(1-0) and $^{13}$CO(1-0)}

Following \citet{Yuan2016} Eqs. 3 and 4, we estimate the $^{13}$CO(1-0) optical depth, $\tau_{13}$, and, multiplying 
by the $^{12/13}$C ratio, the optical depth of the main line, $\tau_{12}$.  This assumes that excitation temperatures 
and surface filling factors (fraction of the beam occupied by emitting molecules) are the same for both isotopologues. 
The $^{13}$CO(1-0) emission is optically thin, with $0.15 < \tau_{13} < 0.32$ calculated in this way, yielding main 
isotope optical depths of order 20.  C$^{18}$O(1-0) emission was not detected in these sources, confirming the low 
optical depths of the $^{13}$CO(1-0) emission.

Assuming a filling factor of unity, the observed brightness temperature  can be used to calculate
the excitation temperature of the $^{12}$CO(1-0) line.  The optical depths obtained previously mean that the main 
line is highly optically thick so that it does not enter in the excitation temperature calculation \citep[see e.g. 
Eq. 88 of][]{Mangum2015} but also that the $^{13}$CO(1-0) column density is almost independent of $\tau_{13}$; but we 
need to assume that the excitation temperatures are the same (or known) for both isotopologues as in \citet{Yuan2016}.
The excitation temperatures vary from 7.3 to 11.4\K\ for the nine sources, apparently reasonable values for the outer 
Galaxy.  Table 3 shows the optical depths, excitation temperatures, CO column densities, and H$_2$ column density 
based on the $^{12}$CO and  $^{13}$CO column density (see Sect. 4.5).  
Table 4 provides the limits to the C$^{18}$O(1-0) emission, obtained by averaging
C$^{18}$O and $^{13}$CO spectra over the maps. 

However, as explained in \citet{Shirley15}, the effective critical density is lower than the standard value for highly 
optically thick lines such as $^{12}$CO(1-0) but not the optically thin $^{13}$CO(1-0).  As a result, we expect that our 
hypothesis of equal excitation temperatures is inappropriate and we test lower excitation temperatures for 
$^{13}$CO(1-0), halfway between the CMB temperature and the $^{12}$CO(1-0) excitation temperature
$T_{ex,13} = T_{bg}+0.5 \, (T_{ex,12}-T_{bg})$, so that the $^{13}$CO(1-0) alternative excitation temperatures 
range from 5 to 7\K.  While ad hoc, this preserves the order among the sources and
remains sufficiently far from $T_{bg}$ that the column density is not close to diverging.  
The results with the modified excitation temperature are also given in Table 3.

One may also wonder whether it is reasonable to assume the same filling factor for the two lines, as we expect 
there to be a low-density region near the edge of the cloud where $^{12}$CO is present (self-shielded) but not 
$^{13}$CO.  This is likely the case but the external radiation field in these extreme outer galaxy sources is very 
low so the difference in filling factors should be small.  Filling factors are extremely structure-dependent and
hence very difficult to estimate, so we have not attempted to change the equal filling factor assumption for 
$^{13}$CO(1-0). 

\subsection{Molecular column densities: HCN(1-0)}
 
For LTE excitation, which is expected given that the energy levels are so close, the optically thin hyperfine 
line ratios should be 5:3:1 (i.e. 0.6 and 0.2 when compared to the main line)
due to the degeneracies of the levels (2F$+$1), such that the total HCN line strength is 1.8 times that of the main component. 
Figure \ref{naive} shows the HFS line ratios expected as a function of the optical depth of the main line.  
Cases of "anomalous" line ratios exist but these {\it all} result in a high $\frac{(0-1)}{(2-1)}$ ratio  
\citep{Kwan75,Guilloteau81} and generally involve high optical depth and/or double-peaked profiles. 
The ($F=1 \rightarrow 1$)/($F=2 \rightarrow 1$) and ($F=0 \rightarrow 1$)/($F=2 \rightarrow 1$) line 
ratios for clouds 21, 56, and 57 are respectively 0.57 and 0.21, 0.60 and 0.18, 0.69 and 0.325.
These values are indicated on Figure \ref{naive}  and indicate that the HCN emission 
from clouds 21 and 56 is optically thin ($\tau \la 0.5$) but cloud 57 has an optical depth $\tau \sim 1$.  
The agreement is excellent between the two line ratios for all three clouds.

 An independent estimate can be obtained  by following  \citet{Yuan2016} who use the 
 CLASS\footnote{\tt https://www.iram.fr/IRAMFR/GILDAS/}  hyperfine structure method to fit the three 
 lines of the summed spectra of clouds 21, 56, and 57.  In addition to velocity and velocity width, this method 
 estimates the optical depth of the main line.  The optical depths obtained are $\tau_{21} = 0.24 \pm 0.15$, 
 $\tau_{56} = 0.66 \pm 0.10$, $\tau_{57} = 1.50 \pm 1.47$ for clouds 21, 56, and 57 respectively, in good agreement 
 with the previous estimates.  

Note that because these are line ratios of the same molecule at essentially the same energy, the optical depths 
are those after taking into account any filling factor, i.e. $not$ beam-averaged.  The ``naive`` and CLASS estimates 
are consistent with optically thin ($\tau<1$) HCN emission from clouds 21 and 56 and $\tau \approx 1$ in cloud 57.

Optically thin means we can estimate the column density from the line intensity.
If we assume that the HCN excitation temperature is that of CO, then the column densities range from 
$3 \times 10^{10}$\scm\ to $2 \times 10^{11}$\scm\ (cloud 57).
If we assume that the HCN excitation temperature is between the CMB temperature and that of CO 
(see previous section for $^{13}$CO), then the column densities are a factor 3 higher, ranging from 
$10^{11}$\scm to $7 \times 10^{11}$\scm\ (cloud 57).
\citet{Yuan2016} assume a very low excitation temperature, where T$_{ex} = 2 h B/k$ (the first energy level,  
4.25\K\ for HCN), and this yields much higher HCN column densities.
These column densities are beam-averaged because they use the integrated intensities, 
unlike the optical depths calculated earlier from the HFS ratios. 
The column densities for the three excitation temperatures are provided in Table 3 but our fiducial value 
is the intermediate one.

\subsection{Molecular column densities: dense gas filling factor}

Because we have an estimate of the optical depth, we can estimate the filling factor by re-arranging 
Eq. 2 of \citet{Yuan2016} as follows
\begin{equation}
T_r = \frac{h\nu}{k} [J(T_{ex})-J(T_{bg})] \, \tau \, f 
\end{equation}
 where $J(T)=[exp(h\nu/kT)-1]^{-1}$ which becomes
\begin{equation}
f = \frac{k \, T_r }{h \nu \tau} \, [J(T_{ex})-J(T_{bg})]^{-1}   
\end{equation}
We use the optically thin approximation $(1-exp(-\tau))\approx \tau$ but at the moderate optical depths 
present here this has little effect. 
We can then use the values of $\tau $ to deduce a limit for the filling factor $f$ for the two $T_{ex}$ above.

In order to apply this reasoning to the other sources, we can average
 the 3 optical depths obtained from the CLASS fits.  We obtain $<\tau> \approx 0.53$ when weighting 
either by the inverse of the uncertainty or by the square of the inverse of the uncertainty.  We apply this to the weaker 
sources which were not mapped, hence probably over-estimating $\tau$.
The filling factors obtained range from 4-16\% for our fiducial HCN(1-0) excitation temperature, see Table 3.
\citet{Wang2020} found dense gas filling factors of 6-28\% in the molecular filament GMF54 \citep{Ragan2014} 
over the regions where $^{13}$CO was detected (equally our case), so the values we find appear reasonable.
GMF54 is beyond the molecular ring and few appropriate comparison sources are available.
We then expect the true (not beam-averaged) column densities to be a factor $1/f$ times higher (col. 10 divided by col 14 in Table 3).
This yields HCN column densities for the cores of $\approx 2 \times  10^{12}$\scm\ except for clouds 56 and 57 which may reach 
$10^{13}$\scm. Note that this is for the total line width and not per $\kms$ and assumes that the dense gas occupies 
a relatively small fraction of the beam.  

\subsection{Molecular column densities: HCO$^+$(1-0)}

 \hcop\  is believed to be less abundant than HCN \citep{Godard2010,Watanabe17} and the turbulent linewidths and 
dipole moments are similar such that if HCN is optically thin, then \hcop\ should be as well.
Taking \hcop\ to be optically thin, we can do the same calculations, again assuming our fiducial excitation temperature
midway between $T_{\rm bg}$ and T$_{ex,CO}$ (Sect 4.1).  The column densities are given in Table 3.
We can see that overall the HCN and \hcop\ column densities are similar: $0.4 < \frac{N_{HCN}}{N_{HCO^+}} < 2$ 
(see columns 10 and 13 of Table 3).

Having established that the HCN(1-0) line, and likely the \hcop line, are optically thin or nearly so, we can ask ourselves 
whether the apparently similar HCN and \hcop\ column densities necessarily reflect the true abundance ratio.
The critical density of HCN is about 7 times higher than that of \hcop\ 
\citep[see ][Table 1 or, for \hcop, Fig. 3 extrapolating 
to our column densities]{Shirley15}, potentially resulting in more severely subthermal excitation for the HCN line.
This could reduce the HCN line intensities, from which we estimate the column densities, more than the \hcop, such that 
the true HCN/\hcop\ abundance ratio could be higher.  As it is difficult analytically, the following section will explore this more 
quantitatively via non-LTE calculations.

\subsection{Molecular column densities: H$_2$}

Observations of $^{12}$CO(1-0), $^{13}$CO(1-0), HCN(1-0), and \hcop (1-0) have been presented.  The latter two molecules 
are considered dense gas tracers as their critical densities are about 100 times higher than for CO where $n_{crit} \approx 1000$\ccm
and less when highly optically thick.  We can estimate the average H$_2$ column density from the CO observations by simply 
dividing by the fractional abundance \citep{Pineda2013,Fontani2012}, such that $N(H_2) = N(^{12}CO) / \chi_{^{12}CO}$ or  
$N(H_2) = N(^{13}CO) / \chi_{^{13}CO}$
where the column densities are given in columns 5 and 8 of Table 3.  The $N(H_2)$ values are presented in the last two columns 
of Table 3.  The H$_2$ column densities derived in this way are significantly higher than when using the standard large-scale $\ratioo$ factor
used in extragalactic observations \citep[e.g. ][]{Bolatto13}.  This is expected as the CO emission from the outer galaxy is weak 
compared to the inner galaxy \citep{Sodroski95,Digel94}, presumably due to lower abundances and lower temperatures. 
Extragalactic observations by  \citet{Sandstrom2013} found that the $\ratioo$ ratio did not change significantly 
within the inner disk (out to 0.7 $R_{25}$).
However, the positions observed here are far beyond $R_{25}$\footnote{Estimates of the disk mass scale length of the Galaxy 
range between 2 and 3 kpc \citep{Binney2023,Bovy-Rix2013} and following \citet{Gusev2012} we estimate the blue-band scale length to be about 3.5kpc, yielding $R_{25} \approx 11$kpc assuming a central disk surface brightness of 21.65 mag/\arcsec$^2$ from \citet{Freeman70}.} 
and the ratios we find by dividing the N(H$_2$) from Table 3 
by I$_{CO}$ from Table 1 are still within the scatter of their Figure 4, such that there is no conflict.

Rather standard values of N(H$_2$) are found, of order $10^{22}$H$_2$ \scm, suggesting that the procedure is reasonable
and that outer galaxy, even extreme outer galaxy, molecular clouds have similar masses and column densities as local clouds.  
Taking the maps  in Figures \ref{map21}, \ref{map56}, and \ref{map57} as representative of the size of the clouds, we obtain 
sizes of 10--15 pc.  If the depth is the same, the average density (i.e. for a volume filling factor of unity) is 
$<n> \approx 10^{22}/3 \, 10^{19} \approx 300$\ccm.  This is sufficient to thermalize the highly optically thick $^{12}$CO(1-0) line 
but not the HCN or \hcop.  In order to reach densities where the dense gas tracers are excited, either the cloud would be a thin sheet
with a line-of-sight depth about 1\% of the extent perpendicular to the line of sight, or the HCN and \hcop\ emission is produced by 
dense clumps within the cloud which do not contribute significantly ($>10$\%) to the cloud mass or the $^{12}$CO(1-0) emission.
We will assume the latter.

It is difficult to estimate the H$_2$ column density of the dense clumps as the HCN and \hcop\ abundances are not well known.
HCN sticks to dust and is thus less abundant in cool dense environments.  \citet{Lahuis2000} suggest that the HCN abundance with 
respect to H$_2$ is $\chi_{\rm HCN} \approx 10^{-8}$ and higher in hot environments.  In the outer Galaxy clouds, we are clearly 
not in the "hot environment" case.  
A study of Planck cold cores by \citet{Yuan2016} found very low abundances, roughly $1.5 \times 10^{-10}$ for both HCN and HCO$^+$.  
Although these clouds were generally much closer, with an average distance of 1.3kpc, this work has the physical conditions closest to 
the outer disk observations presented here.
The \citet{Fontani2022} and \citet{Patra2022} sources (massive star-forming regions) are very different (although in the outer disk) with broader lines, outflows, 
strong HCN and \hcop\ emission and even H$^{13}$CN emission; they do not estimate abundances.
\cite{Pirogov1995} find an HCN abundance $\chi_{\rm HCN} > 10^{-10}$ towards a high-latitude cloud.  \cite{Turner1997} estimate 
$\chi_{\rm HCN} > 10^{-9}$ from observations of translucent molecular clouds.
\citet{Watanabe12} and \citet{Watanabe17} find $\chi_{\rm HCN} \approx 10^{-9}$ towards a low-mass class 0 protostar and 
$\chi_{\rm HCN} \approx 2 \times 10^{-9}$ towards W~51 respectively.  In their extragalactic work, \citet{Gao04b} assume 
$\chi_{\rm HCN} = 2 \times 10^{-8}$.  
\citet{Watanabe14} observed two positions in M~51 and find fractional abundances  $\chi_{\rm HCN} \approx 10^{-9}$.  
\citet{Martin06} and \citet{Aladro11} find values within a factor two of $10^{-9}$ for respectively NGC~253 and M~82 
such that the column density of dense gas is:
\begin{equation}
N(H_2)_{\rm dense} = 10^{21} \times \frac{N_{\rm HCN}}{10^{12}} \times \frac{10^{-9}}{\chi_{\rm HCN}} \times \frac{1}{f} {\rm cm}^{-2} 
\end{equation}
where $N_{\rm HCN}$ is the beam averaged value but, because we divide by the filling factor $f$, $N(H_2)_{\rm dense}$ is
the estimated N(H$_2$) of the dense clumps.
Note that this expression keeps the HCN abundance as a free parameter but introduces reasonable values in order to provide an 
illustrative value of the H$_2$ column density.  The range of $\chi_{HCN}$ that we regard as likely in the extreme outer disk environment is 
$10^{-9} \ga \chi_{HCN} \ga 10^{-10}$.

 A beam averaged HCN or \hcop\ column density 
of a few $10^{11}$\scm (Table 3 cols. 10 and 13) corresponds to beam-averaged dense gas column of $N(H_2)_{dense} \approx 10^{21}\cmt$, to be 
compared with $N(H_2) \approx 10^{22}\cmt$ from cols. 15 and 16 of Table 3.  Thus, depending on the abundance, the mass in the clumps dense 
enough to excite the HCN and \hcop\ molecules represents of order 10\% of the mass. 

\begin{table*}
\begin{center}
\begin{tabular}{llllllll}
Cloud & I$_{\rm HCO^+}$ & V$_{\rm HCO^+}$ & fwhm$_{\rm HCO^+}$ &  I$_{\rm HCN}$ & V$_{\rm HCN}$ & fwhm$_{\rm HCN}$ & rms  \\
 & K km s$^{-1}$ & km s$^{-1}$ & km s$^{-1}$  & K km s$^{-1}$  & km s$^{-1}$ & km s$^{-1}$ & mK \\
\hline
  7 &  0.148$\pm$  0.015 & -102.89 $\pm$ 0.07 & 1.484 $\pm$0.181 &  0.022 $\pm$ 0.018 & -102.92  $\pm$0.18 & 0.494 $\pm$0.779 & 17 \\
 14 &  0.133 $\pm$ 0.015 & -101.11$\pm$  0.08 & 1.407$\pm$ 0.178 &  0.023 $\pm$ 0.008 & -101.28  $\pm$0.06 & 0.373$\pm$ 0.135 & 19 \\
 16 &  0.072  $\pm$0.012 &  -99.69 $\pm$ 0.09 & 1.161 $\pm$0.250 &  0.031  $\pm$0.010 &  -99.36  $\pm$0.09 & 0.615 $\pm$0.249 & 18 \\
 18 &  0.117 $\pm$ 0.013 &  -99.40  $\pm$0.09 & 1.616 $\pm$0.192 &  0.046  $\pm$0.010 &  -99.87  $\pm$0.21 & 1.628 $\pm$0.365 & 12 \\
 30 &  0.100 $\pm$ 0.013 & -104.02  $\pm$0.09 & 1.297 $\pm$0.181 &  0.058  $\pm$0.014 & -104.48  $\pm$0.17 & 1.246 $\pm$0.245 & 22 \\
 34 &  0.201  $\pm$0.018 & -107.25  $\pm$0.09 & 2.221 $\pm$0.255 &  0.038  $\pm$0.012 & -107.17  $\pm$0.18 & 1.143 $\pm$0.386 & 17 \\
 21 &  0.553 $\pm$ 0.018 & -101.10  $\pm$0.03 & 1.801 $\pm$0.073 &  0.086  $\pm$0.012 & -101.20  $\pm$0.08 & 1.020 $\pm$0.166 & 19 \\
       sum & $2-1$ & & &  0.067  $\pm$0.006 & -101.35  $\pm$0.05 & 1.276 $\pm$0.136 & 3.4 \\
      sum & $1-1$ & & &  0.038  $\pm$0.009 &  -96.53  $\pm$0.16 & 1.321 $\pm$0.402  & 3.4\\
      sum & $0-1$ & & &  0.014  $\pm$0.008 & -108.32  $\pm$0.24 & 0.827 $\pm$0.624 & 3.4 \\
56 &  0.319  $\pm$0.013 & -103.55  $\pm$0.05 & 2.345 $\pm$0.107 &  0.126  $\pm$0.013 & -103.45  $\pm$0.09 & 1.828 $\pm$0.239 & 12 \\
       sum & $2-1$ & & &  0.096  $\pm$0.009 & -103.57  $\pm$0.08 & 1.794 $\pm$0.189 & 4.3 \\
      sum & $1-1$& & &  0.058  $\pm$0.013 &  -98.73  $\pm$0.19 & 1.744 $\pm$0.448 & 4.3 \\
      sum & $0-1$& & &  0.017  $\pm$0.011 & -110.70  $\pm$0.37 & 1.126 $\pm$0.761 & 4.3 \\
 57 &  0.474  $\pm$0.016 & -102.23  $\pm$0.03 & 2.013 $\pm$0.082 &  0.133  $\pm$0.016 & -102.17  $\pm$0.09 & 1.463 $\pm$0.200 & 16 \\
       sum & $2-1$ & & &  0.083  $\pm$0.009 & -102.37  $\pm$0.10 & 1.799 $\pm$0.236  & 5.9\\
      sum & $1-1$ & & &  0.057  $\pm$0.012 &  -97.38  $\pm$0.19 & 1.821 $\pm$0.446  & 5.9\\
      sum & $0-1$ & & &  0.027  $\pm$0.014 & -109.59  $\pm$0.51 & 1.980 $\pm$1.116 & 5.9\\
 \end{tabular}
\caption[]{Results of observations of central positions, deduced from gaussian fits.  
Intensities are in main beam scale but for HCN are only of the main $F=2 \rightarrow 1$ transition.  
The spectra for clouds 21, 56, and 57 have been summed (see Fig. \ref{hyper} and the lines marked "sum" provide 
the fits to the individual hyperfine components.
The expected velocity differences are 4.842 and $-7.064$ km s$^{-1}$ with respect to the main (2-1) component.  
The satellite lines were fit using as initial guess (for the fit algorithm of CLASS) the main line velocity $+$ 4.842 or -7.064 
for the $1-1$ and $0-1$ transitions respectively.
That the velocities do not show exactly these differences provides an independent estimate of the uncertainty in the velocity determination.
Rms noise levels are in mK converted to main beam scale and per 0.2 km s$^{-1}$ channel.}
\end{center}
\end{table*}

\begin{table*}
\begin{center}
\begin{tabular}{llllllllllllllll}
Cl & T$_{\rm CO}$ & $\chi_{\rm 12/13}$ & T$_{\rm ex,CO}$ &  N$_{\rm CO}^a$ & N$_{\rm ^{13}CO}^a$ & 
T$_{\rm ex,^{13}CO}^b$ & N$_{\rm ^{13}CO}^b$ & N$_{\rm HCN}^a$ & N$_{\rm HCN}^b$ & N$_{\rm HCN}^c$ & N$_{\rm HCO^+}^a$ & 
N$_{\rm HCO^+}^b$ & f$^b$ & N(H$_2$)$^a$ &  N(H$_2$)$^b$ \\
 &  &  &  & $10^{17}$  & $10^{15}$ &  & $10^{15}$ & $10^{11}$ & $10^{11}$ & $10^{11}$ & $10^{11}$ & $10^{11}$ & \% & log & log \\
 & K &  & K & \scm\  & \scm\ & K & \scm\ & \scm\ & \scm\ & \scm\ & \scm\ && &  &  \\
 7 &   5.4 &  93.20 &  8.75 &  1.82 &  1.17 &  5.74 &  4.23 &  0.29 &  0.91 &  2.91 &  0.63 &  2.01 &   5.8 & 21.67 & 22.00 \\
14 &   4.5 &  93.67 &  7.80 &  2.83 &  1.71 &  5.27 &  6.33 &  0.40 &  1.28 &  3.04 &  0.75 &  2.43 &   7.2 & 21.87 & 22.19 \\
16 &   4.0 &  93.20 &  7.30 &  2.52 &  1.74 &  5.02 &  6.48 &  0.63 &  2.07 &  4.10 &  0.48 &  1.57 &  10.9 & 21.81 & 22.19 \\
18 &   5.4 &  92.73 &  8.75 &  1.96 &  1.35 &  5.74 &  4.87 &  0.60 &  1.91 &  6.09 &  0.50 &  1.59 &  12.1 & 21.69 & 22.06 \\
30 &   5.0 & 102.13 &  8.35 &  0.58 &  1.61 &  5.54 &  5.88 &  0.84 &  2.71 &  7.68 &  0.47 &  1.53 &  16.4 & 21.28 & 22.29 \\
34 &   5.8 & 105.42 &  9.15 &  2.08 &  1.24 &  5.94 &  4.45 &  0.45 &  1.41 &  5.03 &  0.77 &  2.44 &   9.3 & 21.87 & 22.23 \\
21 &   8.0 &  95.55 & 11.40 &  0.96 &  0.89 &  7.06 &  3.04 &  0.55 &  1.66 &  9.84 &  1.32 &  4.03 &  12.3 & 21.42 & 21.90 \\
56 &   6.0 & 127.04 &  9.35 &  3.14 &  1.91 &  6.04 &  6.81 &  1.45 &  4.58 & 17.18 &  1.16 &  3.68 &   8.6 & 22.31 & 22.75 \\
57 &   5.5 & 125.63 &  8.85 &  2.98 &  1.16 &  5.79 &  4.18 &  2.27 &  7.23 & 23.72 &  1.96 &  6.26 &   4.0 & 22.27 & 22.52 \\
\hline
 \end{tabular}
\caption[]{Temperatures, column densities, and filling factors as described in Section 4.  Col. 2 is the observed CO line 
temperature, col. 3 the radially varying $^{12/13}$CO abundance ratio, col. 4 is the CO excitation temperature, and columns 5 and 6 are the 
deduced CO and $^{13}$CO column densities, all following Sect. 3.2 of \citet{Yuan2016}.  
Col. 7 is our fiducial excitation temperature for the optically thin or close to thin species (see Sect. 4.1).
Columns 8 - 13 are the column densities derived using the excitation temperatures from Sect. 4.1 and \citet{Yuan2016}, see footnotes below.
Col. 14 is the estimated filling factor from Sect. 4.3.
The H$_2$ column densities in the last two columns are derived as in the first paragraph of Sect. 4.5, where col 16. is the H$_2$ column deduced from the $^{13}$CO line assuming an excitation temperature as in ($b$) below.
($a$) Assumes all excitation temperatures are those derived for CO (col. 4).
($b$) Our fiducial case which attempts to account for subthermal excitation temperatures such that T$_{\rm ex} = T_{bg}+0.5 \times (T_{\rm ex,CO}-T_{bg})$.
($c$) Follows \citet{Yuan2016} hypothesis where HCN and \hcop\ excitation temperatures are equal to $2 h B/k$.}
\end{center}
\end{table*}

\begin{table}
\begin{center}
\begin{tabular}{lllll}
Cloud & $^{13}$CO & C$^{18}$O & ratio & $\chi_{13/18}$ \\
 & K km s$^{-1}$  & K km s$^{-1}$  & & \\
21  &  $ 0.60 \pm 0.06$  &  $0.03 \pm 0.06$  & $\ga20 \pm 10$ & 9.6 \\
 56  & $1.39 \pm 0.07$ &   $0.20 \pm 0.07$ & $\ga7.0\pm 2$ & 10.3  \\
 57  &   $0.85 \pm 0.06$   & $0.11 \pm 0.06$ & $\ga7.7 \pm 3$ & 10.3\\
\hline
 \end{tabular}
\caption[]{$^{13}$CO(1-0) and C$^{18}$O(1-0) intensities averaged over the maps.
$\chi_{13/18}$ is the $^{13}$CO to C$^{18}$O abundance ratio at the galactocentric distance of the cloud, which for optically thin emission and equal excitation temperatures would be the intensity ratio. The last two columns can legitimately be compared.}
\end{center}
\end{table}

\section {Physical conditions via radiative transfer calculations}

The analytical calculations in the previous section are necessarily based on the hypothesis that the molecules are in 
Local Thermal Equilibrium.  LTE is a strong assumption which is probably not valid, particularly for  high-dipole 
moment molecules like HCN and \hcop.  We made a first attempt to account for this by allowing for a lower excitation temperature
for the optically thin (or nearly so) molecular emission.  In this section we make non-LTE calculations in order to 
go a step further and in particular to treat the HCN and \hcop\ molecules separately because although they both have high 
dipole moments, their critical densities are different and the non-LTE transfer calculations enable us to interpret the 
factor $\sim 2$ difference in intensity. 

\subsection{Model setup}
To constrain the volume and column densities of our observed outer Galaxy clouds, we have run a grid of non-LTE models of the HCN and \hcop\ line emission using the radiative transfer program RADEX\footnote{\tt https://personal.sron.nl/$\sim$vdtak/radex/index.shtml} \citep{vandertak07}.
This program solves for the relative populations of the rotational energy levels of the molecules, taking collisional and radiative (de-)excitation into account, and treats the line radiative transfer with an escape probability formalism.
The free parameters in the calculation are the kinetic temperature, the gas density, and the molecular column density.
In the calculations, we assume a background temperature of 2.73~K, a turbulent line width of $1.0~\kms$, and a uniform medium without velocity gradients. The linewidths of these clouds are slightly above $1.0~\kms$ and any column density per turbulent linewidth should be multiplied by the linewidth (Table 2) to obtain the total column density.

\citet{Watanabe17} indicate that $\chi_{\rm HCO^+} \approx \chi_{\rm HCN} /$3 on average in the sources  they consider (their Figures 10, 11), 
with $\chi_{\rm HCO^+} \la 10^{-9}$.  Although somewhat uncertain, we take this abundance ratio for our calculations.  The decreasing [N/O] 
ratio would favor \hcop\ in the outer disk but \hcop\ requires a source of ionization. The competition between these effects is difficult to evaluate 
and, to our knowledge,  model calculations have not provided insight to this.  Based on the estimates in \citet{Yuan2016}, we see no evidence 
for lower \hcop/HCN abundance ratios.

Spectroscopic data for the HCN and \hcop\ rotational and hyperfine structure were taken from the Cologne Database for Molecular Spectroscopy (CDMS)\footnote{\tt http://cdms.de} \citep{mueller2005}.
The \hcop\ calculations use collision data with \hh\ by \citet{flower1999}, which cover rotational levels up to $J$=20 and temperatures of 40--200\,K. 
For the collisional (de)excitation of the lowest rotation-hyperfine levels of HCN with \hh, two calculations exist, which both cover our relevant energy levels (up to $J$=8--10) and temperatures (down to 5\,K and up to 30--100\,K), but which differ in the ratio between the rates for radiative (\DJ=1) and non-radiative (\DJ=0, \DF=0,$\mp$1) transitions. 
The calculation by \citet{benabdallah2012} used a potential energy surface for the interaction with a reduced dimensionality, averaging over \hh\ orientations, and found that collisional deexcitation through non-radiative (cross-$F$) channels is $\approx$10$\times$ faster than through \DJ=1 channels.
However, after upgrading the close coupling calculations to full dimensionality, \citet{hernandez2017} found that the collisional deexcitation through both types of channels proceeds at approximately equal rates. 
For comparison, the quasi-classical calculation by \citet{green1974} did not resolve the hyperfine structure of HCN and used He as collision partner. The corresponding datafile on the Leiden Atomic and Molecular Database (LAMDA) \citep{schoeier2005}\footnote{\tt http://home.strw.leidenuniv.nl/$\sim$moldata/} assumes that the radiative transitions have equal collision rates, while the cross-$F$ rates are zero.
For lack of experimental data on the relative propensities of cross-$F$ rates, we have explored all 3 sets of collision rates in our models for HCN.

\begin{figure}
\centering
\includegraphics[width=\hsize{},angle=0]{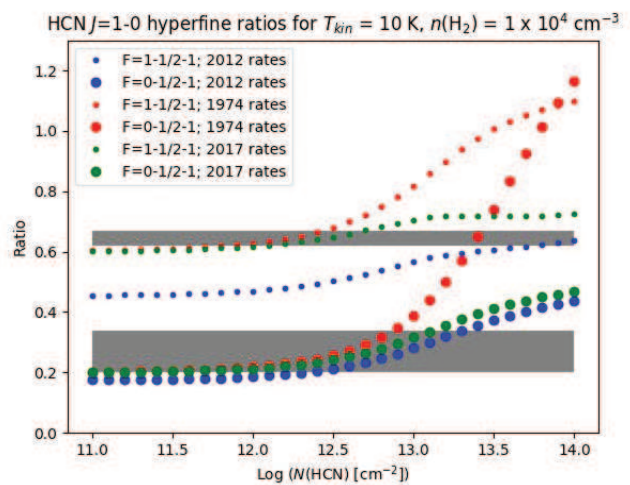}
\caption{Intensity ratios of the hyperfine components of the HCN $J$=1--0 line, calculated with RADEX as a function of $N$(HCN). Grey areas indicate observed ranges for the $F$=1--1/2--1 (top) and 0--1/2--1 (bottom) ratios, respectively 0.6, 0.6, 0.7 and 0.2, 0.2, 0.3 for clouds 21, 56, and 57. The numbers indicating the rates refer to the years of publication of the rates.
}
\label{f:hfs}
\end{figure} 

\subsection{Model results}
To constrain the column densities of the clouds, we model the intensity ratios of the hyperfine components of the HCN $J$=1--0 transition. 
Figure~\ref{f:hfs} shows the result for an assumed \tkin\ of 10\,K and $n$(\hh) = 10$^4$\,\ccm, and 
calculations for \tkin\ in the 10--30\,K range and $n$(\hh) in the 10$^3$--\pow{3}{4}\,\ccm\ range indicate that our results depend only weakly on the assumed gas density and kinetic temperature.
Column densities of $N$(HCN) $\approx$\pow{3}{12}\,\scm\ are seen to be consistent with both observed ratios (indicated by the grey areas), but only if the  \citet{hernandez2017} collision data with the similar cross-$F$ and cross-$J$ rates are used.
Using the data with the large cross-$F$ rates, there is no column density of HCN which matches both observed ratios simultaneously.
The collision data with the cross-$F$ rates equal to zero are also able to match the observed hyperfine ratios, but the inferred $N$(HCN) is less reliable since these hyperfine collision rates were assumed rather than calculated. These rates also lead to unrealistically high hyperfine ratios toward high HCN column densities.
The models also reproduce the observed absolute intensities of the HCN lines, for similar column densities as indicated by the hyperfine ratios. 
We conclude that the HCN line emission indicates $N$(HCN) $\approx$\pow{3}{12}\,\scm, using the collision data by \citet{hernandez2017}.  

\begin{figure}
\centering
\includegraphics[width=\hsize{},angle=0]{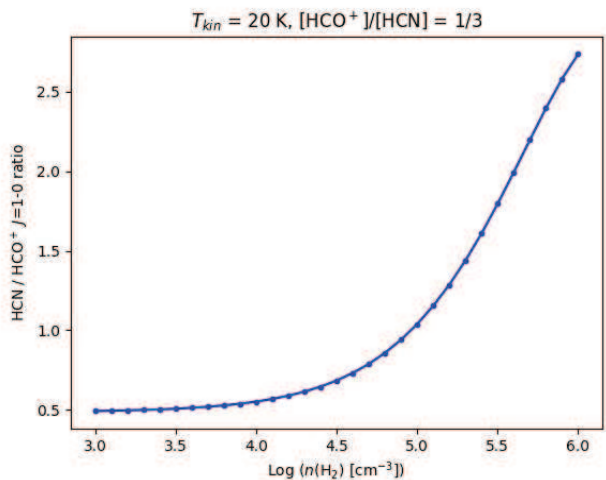}
\caption{Intensity ratio of the HCN and \hcop\ $J$=1--0 lines, calculated with RADEX as a function of $n$(\hh).}
\label{f:hcop}
\end{figure} 

To constrain the volume densities of the clouds, we model the ratios of the HCN and \hcop\ $J$=1--0 lines. 
These calculations assume \tkin\ = 20\,K and an abundance ratio of [HCN]/[\hcop] = 3 \citep{Watanabe17}. 
Figure~\ref{f:hcop} shows the result as calculated with RADEX using the HCN collision rates by \citet{hernandez2017}.
Our observed line ratio of $\approx$1/2 (summed over the HCN hyperfine components) is seen to correspond to \hh\ densities in the $10^3$--$10^4$\,\ccm\ range.
The analytical calculations in the previous section yielded similar HCN and \hcop\ columns but the degree of subthermal 
excitation could not be assessed.  Nonetheless, if $\chi_{\rm HCO^+} > \chi_{\rm HCN}/3$, then a higher density would be 
required to yield the observed HCN/\hcop\ line ratio. 
For example, if $\chi_{\rm HCO^+}/ \chi_{\rm HCN} = 0.5$ (and not 1/3), then we would deduce a density of  $n$(\hh) $\approx 10^{4.5}\cc$.

\subsection{HCN and HCO$^+$ as dense gas tracers}

Depending on the HCN and \hcop\ abundances, the H$_2$ column density deduced from HCN could be comparable to that deduced from the CO observations (Table 3).  {\it Can the gas be the same?}  We can test this using the intensities of the lines. Assuming that the majority of the CO emission comes from gas with a density $n \la 3000 \cc$ (see discussion in Sect. 4.5), RADEX calculations show that for $n \la 3000\cc$, $I_{HCN}/I_{CO} \la 0.001$ 
(H$_2$ column assumed is $10^{21}\cmt$ with abundances $\chi_{CO}=10^{-5}$ and $\chi_{HCN}=10^{-9}$).  
Since $I_{HCN}/I_{CO} \ga 0.01$ from Tables 1 and 2, the HCN emission does not come from the gas emitting the majority 
of the CO emission but rather from clumps or cores within the less dense ambient molecular gas.
n principle, the CO could come from gas with $n > 3000 \cc$.  However, not only would the clouds necessarily be much thinner along the line of sight 
but also the $^{13}$CO emission would be quite unusual, much stronger in the higher$-J$ transitions than in $^{13}$CO(1--0).  
The few data available for the $^{13}$CO(2--1) line, or even CO(2--1), at large scales in the outer parts of galaxies suggest that the 
higher-$J$ transitions are not stronger \citep{Sakamoto97,Braine_n4414a}. 
Hence, in these outer disk clouds, HCN and HCO$^+$ trace dense gas ($n > 10^3 \cc$).

Figure~\ref{hcn_area} shows that the HCO$^+$ line is on average twice as strong as the HCN line.
To make this figure, the HCN(1--0) main line intensities were multiplied by 9/5 to account for the flux in the weaker satellite lines which have low signal-to-noise ratios and thus can add considerable noise to the sum of the intensities.  As seen in Sect. 4, the optical depth is low so this procedure should be appropriate.  Since both lines are optically thin, we have used RADEX to estimate the typical density at which the HCO$^+$(1--0) is twice the total HCN(1--0) line intensity.  For a column density ratio of 3, this occurs at $n \la 10^4 \cc$ (T$_{\rm kin}=20$K).   
If in fact the HCO$^+$ abundance is higher (than 1/3 of the HCN abundance), then the density implied by the observed flux ratio is higher.
A larger fraction of the HCO$^+$ emission, compared to HCN, comes from the lower density CO-emitting gas and this is why the above calculation sets a lower limit to the typical density of the dense component.  This is also seen in Fig.~11 of the observations of W3(OH) by \citet{Nishimura17} which shows that the HCN(1--0)/HCO$^+$(1--0) ratio decreases towards the more extended and more diffuse material.

We obtain a typical clump or core size of $L \approx N/n$ where
$N$ is the dense gas column density defined in Eq. 3 such that 
\begin{equation}
M_{cl} = \rho V \approx n 2 m_p f_h^{-1} (N/n)^3 
\end{equation}
where $2 m_p f_h^{-1} $ is the average molecular mass.  Adopting reference values of $N_{HCN} = $ \pow{3}{11}\,\scm, $n = 10^4\cc$, 
$\chi_{HCN} = 10^{-9}$, and $f = 0.1$, we can express the clump or core mass as follows by substituting Eq. 3 into Eq. 4 
\begin{equation}
\begin{split}
M_{cl} & = \frac{2 m_p}{f_h} 10^{-8} \left( \frac{10^4}{n}\right) ^2 (3\times10^{21})^3 \left( \frac{N_{HCN}}{3 \times 10^{11}}\right) ^3 \left( \frac{10^{-9}}{\chi_{HCN}} \right) ^3 \left( \frac{0.1}{f} \right) ^3 \\
 & = 0.62 \left( \frac{10^4}{n}\right) ^2 \left( \frac{N_{HCN}}{3 \times 10^{11}}\right) ^3 \left( \frac{10^{-9}}{\chi_{HCN}} \right) ^3 \left( \frac{0.1}{f} \right) ^3 M_\odot
\end{split}
\end{equation}
with a hydrogen fraction of $f_h=0.73$.
The mass in such a clump is of order a solar mass but highly dependent on the abundance, 
such that if $\chi_{HCN} = 10^{-10}$ then the clump mass becomes $620 M_{\odot}$.  
The other parameters are constrained by observations: $N_{HCN}$ by the HCN flux, $N_{HCN}/f$ 
by the HCN optical depth limit, and $n$ by the HCN/\hcop\ flux ratio.
 
\section{Link between star formation and dense gas fraction}

\begin{figure}
\centering
\includegraphics[width=0.9\hsize{},angle=0]{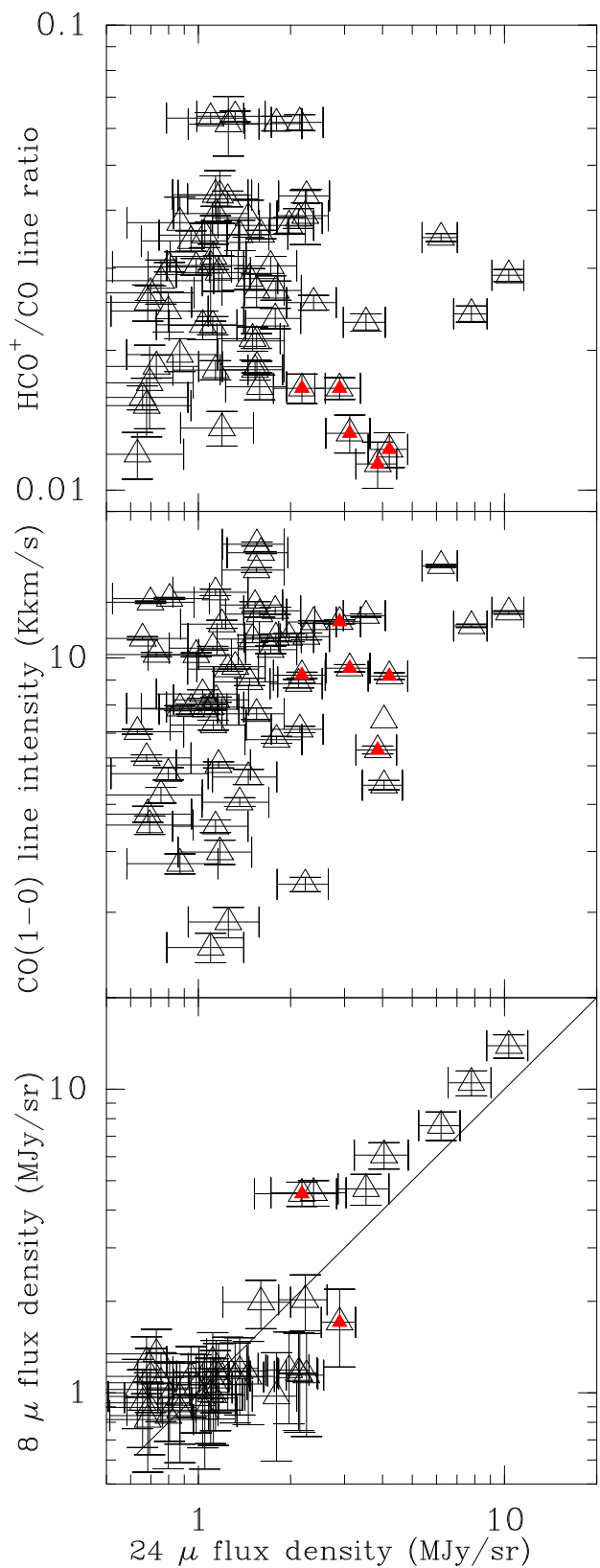}
\caption{Comparison of molecular line intensities and tracers of star formation.  The bottom panel 
compares the SFR tracers: background-subtracted 24 $\mu$m emission and
 8 $\mu$m emission, also background-subtracted (the background emission is considerably 
stronger than any stellar contribution at 8 $\mu$m). The middle panel shows how the $^{12}$CO 
intensity varies with SFR. The top panel shows how the dense gas fraction, as traced 
by the \hcop/CO ratio, varies with SFR.  The red triangles indicate the sources which 
are at high galactic latitude. It should be immediately apparent that these sources are not 
distinguishable from the others in the lower two frames but occupy a specific region with 
low dense gas fraction in the top panel. Neither 8 nor 24 $\mu$m data are available for cloud 30.}
\label{f:sfr-gas}
\end{figure} 

\begin{figure}
\centering
\includegraphics[width=0.9\hsize{},angle=0]{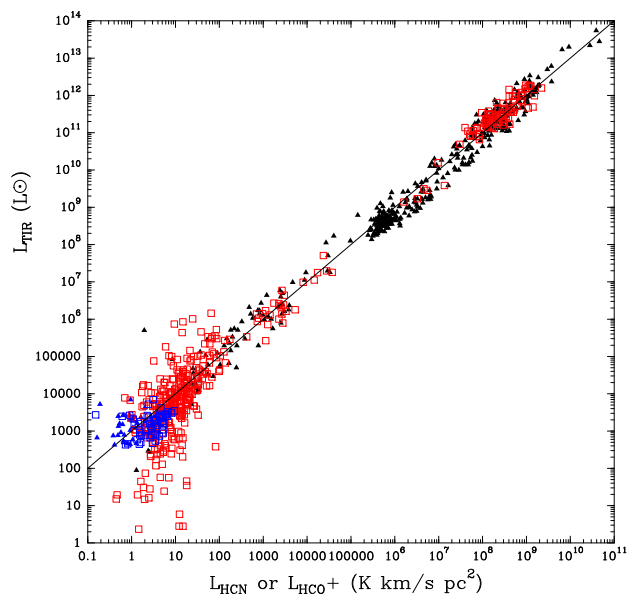}
\caption{Link between dense gas mass (HCN or \hcop\ luminosity on the x-axis) and star formation rate 
(IR luminosity on the y-axis).  The line shows a linear relation with 
$L_{FIR}/L_{HCN}$ or $L_{FIR}/L_{HCO^+} = 1000 L_\odot (\K \kms {\mathrm pc}^2)^{-1}$.  
HCN data are shown as filled triangles and \hcop\ as open squares.
The outer disk data presented in this paper are shown in blue. 
The sources for the data are: \citet{Chin97}, \citet{Chin98}, \citet{Gao04}, \citet{Gao07}, \citet{Brouillet05}, \citet{Wu2005}, 
\citet{Ma2013}, \citet{Buchbenderthesis}, \citet{Krips2008}, \citet{Juneau2009}, \citet{Privon2015}, \citet{Gracia-Carpio08}, 
\citet{Garcia-Burillo2012}, \citet{Chen2015}, \citet{Chen2017}, \citet{Usero2015}.
}
\label{f:sfr-gas2}
\end{figure} 

\citet{Gao04} suggested that the star formation rate is proportional to the mass of dense gas such that starburst galaxies have high 
dense gas masses. In ultraluminous IR Galaxies (ULIRGs), there is generally a huge amount of gas concentrated near the center and hence 
the dense gas mass (HCN flux) and fraction (typically the HCN/CO flux ratio) are high.
Here we examine smaller scales so, to compare dense gas with star formation, we have compared these line ratios for 
positions with and without a class I or II source.  The HCN/CO and HCO$^+$/CO flux ratios are marginally higher for the positions with a 
class I or class II source but not significantly.

Aside from the good correlation between 8 $\mu$m and 24 $\mu$m emission, the main result is that the HCN/CO and HCO$^+$/CO ratios are a factor two lower for the sources at high galactic latitude ($b > 2^\circ$, sources 7, 14, 16, 18, 30, and 34, see top panel of Fig. \ref{f:sfr-gas}).  These sources are identified by red triangles in Fig. \ref{f:sfr-gas}.  

In a spiral galaxy, the molecular-to-atomic gas ratio at kpc scales is largely controlled by the local hydrostatic 
pressure \citep[see][]{Blitz04}, itself a function of the stellar and gaseous surface densities.  
As the stellar surface density decreases with galactocentric distance (spiral disks are approximately exponential), 
so does the molecular fraction. The same logic applies to the dense gas fraction which is presumably why 
the HCN/CO and \hcop/CO flux ratios we find are below the inner disk ratios \citep[e.g.][]{Usero2015}.
It likely also explains why the high-latitude clouds (200pc or more above the galactic mid-plane) of our sample or from \citet{Yuan2016} have 
lower HCN/CO and \hcop/CO flux ratios.

Far-IR emission comes mainly from dust heated by young stars and as such is a standard tracer of star formation.
There is a clear correlation between the presence of star formation, as traced by Far-IR emission, and molecular line emission at large scales.  
We will use the Spitzer 24 $\mu$m and 8 $\mu$m emission to trace the SFR because our sources are unobscured and the Spitzer data have high angular resolution, respectively 6\arcsec\ and 2\arcsec.  The 24 $\mu$m emission is a standard tracer of star formation \citep{Calzetti07} but the 8 $\mu$m emission varies strongly as a function of environment and particularly metallicity \citep{Draine07}.  However, for a given environment (outer disk in the present case), it can be used to trace the SFR \citep{Boquien15}. 

The 24 $\mu$m and 8 $\mu$m emission\footnote{https://sha.ipac.caltech.edu/applications/Spitzer/SHA} 
were background subtracted by examining a histogram of the values in the maps.  
The value considered to be ''background'' is where the number of pixels reaches about 1\% of the maximum number of pixels in 
the histogram.  Thus we are not sensitive to variations in a small number of pixels yet we use low values in the map.  
The 8 and 24 micron measurements are continuum, meaning that they see all matter along the line of sight and, 
unlike for the spectra, outer galaxy and more local material cannot be distinguished.  As we are looking in the plane of the galaxy, 
there is emission virtually everywhere such that the background value is not well defined. 
The 8 and 24$\mu$m fluxes provided can only be considered best estimates.
Towards our clouds, with the exception of cloud 30, the outer galaxy cloud is the main source (this is deduced from the relative CO line strength at each velocity).  However, along other lines of sight within the 8 and 24 $\mu$m maps, this is not necessarily the case so we subtract a low value rather than the most common value (which reflects the emission and not the background).  We make sure not to take extremely low values which could be influenced by noise.  
Figure \ref{f:sfr-gas} shows that the 8 and 24$\mu$m fluxes are highly correlated, which would not be the case if the estimated background fluxes were highly incorrect.  The fluxes were measured with two different instruments (IRAC and MIPS on the Spitzer satellite), making a coincidence even less likely. 

Figure \ref{f:sfr-gas} shows how  the molecular line intensity varies as a function of the SFR.
Not all sources have both 24 $\mu$m and 8 $\mu$m available so, after showing that the 24 $\mu$m and 8 $\mu$m fluxes are highly correlated ($r=0.82$, bottom panel), we trace the SFR with the 24 $\mu$m fluxes but use 8 $\mu$m fluxes when no 24 $\mu$m data are available.  The CO intensities are weakly correlated with SFR ($r \approx 0.2$, middle panel) and the HCN and \hcop intensities (not shown in Fig. \ref{f:sfr-gas}) are not significantly correlated with SFR.  The \hcop/CO ratio (top panel), tracing the dense gas fraction, is not significantly correlated with SFR but the sources further than 300 pc above the galactic plane stand out as having very low \hcop\ (and HCN, not shown) emission.  These sources do not stand out in other ways.

Comparing to larger scales, the dense gas tracers \citep[or dense gas mass following][]{Gao04}, and star formation rate, as calculated from the 8 or 24 $\mu$m flux \citep{Boquien10b, Galametz2013}, follow the extrapolation of the trends found in \citet{Chen2015} and \citet{Shimajiri2017}, as shown in Fig. \ref{f:sfr-gas2}. 
The reason for the absence of a correlation between L$_{HCN}$ and SFR is likely the similarity of the regions we observe, all occupying the same area of Fig. \ref{f:sfr-gas2}.

\section{Comparison with other environments}

In the extreme outer Galaxy clouds observed here, the line intensity ratios are approximately: CO/HCN$\sim 70$, CO/HCO$^+ \sim 35$, and CO/$^{13}$CO$\sim7.5$.
Thus, the CO emission is stronger relative to the other lines than in the majority of sources observed thus far  \citep[see e.g. Fig. 8 of][]{Chen2015}.

However, in the nearly square degree region of Orion B mapped by \citet{Gratier17b}, they obtain very similar ratios for their median values of CO, $^{13}$CO, HCO$^+$, and HCN, although the HCN emission is stronger after correction for flux in the satellite lines.  The median values tend to sample the more extended lower column density regions, as also found recently by \citet{Evans2020}.

The \citet{Shimajiri2017} work on nearby molecular clouds showed that HCN and HCO$^+$ emission often traced star formation while the H$^{13}$CN and H$^{13}$CO$^+$ lines were sensitive to the total H$_2$ column.  These regions are much denser than the outer Galaxy clouds studied here and the HCN and HCO$^+$ emission is highly optically thick. The net effect is that the intensity of HCN and HCO$^+$ is highly dependent on gas temperature (thus star formation) rather than density.  In the highest column density regions, the main isotope lines are strongly self-absorbed. 
HCN emission in outer Milky Way clouds is optically thin and does not trace star formation and so it appears akin to the H$^{13}$CN in \citet{Shimajiri2017}, which traces column density in high volume density regions. While \citet{Shimajiri2017} do not calculate H$^{13}$CN abundances, the H$^{13}$CO$^+$ abundances are within a rather narrow range:$1.5-5.8 \times 10^{-11}$ with respect to H$_2$.  These values are consistent with $\chi_{HCO+} \sim 10^{-9}$ and the column densities ($N_{H^{13}CO^+} \sim 10^{12}\cmt$) are similar to our estimates for the main isotope.

\citet{Yuan2016} present HCN and HCO$^+$ observations, also with the PMO telescope, of cool clumps identified by the Planck satellite. They are at approximately the solar circle (mean distance of 1.3kpc), neither outer nor inner Milky Way.  Their estimated column densities are variable but have median N$_{\rm HCN}$ and N$_{\rm HCO^+} \approx 10^{12}\cmt$, although the ratio is quite variable.  Their abundance estimates differ from source to source but are mostly  $\chi \sim 10^{-10}$ for both HCN and HCO$^+$. Other Galactic works include \citet{Park1999} and \citet{Barnes2011} who assume higher abundances, typically $\chi \sim 10^{-9}$.  In the cool clouds, the HCO$^+$ line is stronger than HCN.

So far we have compared with other Galactic observations but extragalactic sources allow for a much greater range in properties.  A relatively distant region in M~51 was observed with the IRAM interferometer by \citet{Chen2017} who found HCN/CO ratios similar to ours but generally higher HCN/HCO$^+$ ratios.  M~31 was observed by \citet{Brouillet05} and, in the outer parts, a similarly low HCN/CO ratio was found but again a higher  
HCN/HCO$^+$ ratio.  
The Large Magellanic Cloud (LMC) provides an environment with a lower metallicity and a higher radiation field than our observations of the outer Milky Way but at similar scales.  \citet{Chin97} observed several lines in the LMC and found again that the HCO$^+$ emission was stronger than the HCN (or HNC), unlike large Milky Way molecular clouds or star-forming galaxies (their Table 10).
Subsequently, \citet{Nishimura2016b}, \citet{Anderson2014}, \citet{Galametz2020}, and \citet{Seale2012} observed CO-strong molecular clouds in the LMC and consistently confirm the weak emission from Nitrogen-bearing molecules, as well as strong CCH emission.  
\citet{Braine17} observed the low-metallicity galaxies M~33, IC~10, and NGC~6822 in the local group and found ratios quite similar to those presented here. These authors concluded that the low HCN/HCO$^+$ ratio was due to weak HCN emission rather than particularly strong HCO$^+$ emission.  This was attributed to the decrease in N/O with metal deficiency.

The low HCN/HCO$^+$ ratios we observe in the extreme outer Galaxy are probably due to a combination of low 
volume density and subsolar metallicity.  A gradient in N/O exists in the Galaxy \citep{Magrini2018}.  
However, while \citet{Arellano2021} find a N/O gradient of $-$0.015 dex kpc$^{-1}$ from optical observations, 
\citet{Rudolph2006} find a N/O gradient of $-$0.03 dex kpc$^{-1}$ from Far-IR spectral line data.  
As a result, the N/O ratio for the distant clouds should be approximately half of the solar circle value.  
If the new values underestimate the gradient, then the low HCN fluxes could be entirely due to a decrease in abundance.  
Two features lead us to believe that volume density is also an issue.  First of all, the new gradient uses high-quality 
GAIA DR2 distances.  Secondly, the low molecular fraction in and the diffuse nature of the outer parts of 
spiral galaxies argues for a lower hydrostatic pressure as compared to the inner disk \citep[see e.g.][]{Blitz04}.  
While the molecular line ratios are not clearly related to the level of star formation, there is a clear difference between the 
clouds in the plane and those at $b>2^\circ$ in that the HCO$^+$/CO ratio is lower for the latter 
(Fig. \ref{f:sfr-gas}, but also \hcop/$^{13}$CO and the same ratios with HCN). 
Our clouds at higher latitude are at $h \ga 300$pc from the galactic 
plane such that the local pressure should be lower.  

We find a similar effect in the \citet{Yuan2016} sample.
Clouds with $h>180$pc above the plane have significantly lower average and median I$_{HCO^+}$/N($^{13}$CO) 
and I$_{HCN}$/N($^{13}$CO) values than the clouds at $h<180$pc. The mean values are 50\% higher in the plane 
and the median is twice as high for both I$_{HCO^+}$/N($^{13}$CO) and I$_{HCN}$/N($^{13}$CO).  N($^{13}$CO)
has been used as the reference column density but the higher dense fraction (stronger HCN and \hcop\ emission) 
close to the Galactic plane is equally found using the $^{12}$CO column density \citep[see Table 3 of ][]{Yuan2016}.

\section{Conclusions}

The dense gas tracers HCN(1-0) and \hcop(1-0) were observed in a sample of extreme outer Galaxy clouds 
($15\la R_{gal} \la 21$kpc) from the \citet{Sun15} unbiased CO survey.  The clouds were chosen for their 
reasonably strong CO ($I_{CO}>7 \kms$) and $^{13}$CO ($I_{^{13}CO}>1.4 \kms$) emission.  
None were detected in C$^{18}$O(1-0) but the limits are close to what would be expected for optically thin emission
so far out in the disk.

1.  The HCN and \hcop\ emission is weak, both in an absolute sense and relative to the CO lines (Tables 1 and 2).  
$I_{\rm HCO^+} \approx 0.03 I_{\rm CO} \approx 0.21 I_{^{13}{\rm CO}} \approx 2 I_{\rm HCN}$ after accounting for 
the satellite HCN lines.  These ratios are typical of a subsolar metallicity environment, where N is more affected than O.

2. The $^{12}$CO(1-0) emission is highly optically thick but the $^{13}$CO(1-0) is optically thin, $\tau_{13}\la 1$.
The estimated H$_2$ column densities are $\sim 10^{22}$\scm, depending on the hypotheses (Table 3), but in all cases 
several times higher than would be deduced from a standard extragalactic $\ratioo \sim 2 \times 10^{20}\Xunit$ conversion 
factor \citep[e.g.][]{Bolatto13}.

3.  The lines are narrow and the hyperfine ratios of HCN indicate that the emission is optically thin, $\tau_{HCN(1-0)}\la 1$.
We obtain beam-averaged HCN column densities, 
for our fiducial excitation temperature, of $1 - 7 \times 10^{11}$\scm\ (Table 3, col 10).
With the optical depths, we can estimate the dense gas filling factor, 
again dependent on the assumed $T_{ex}$, and we obtain values of $10\pm 6$\% (Table 3), in reasonable agreement 
with other estimates. Other molecules and transitions can trace gas substantially denser, which would then have
a lower filling factor.  Similar results are obtained for the \hcop\ line (see Table 3).  

4.  The H$_2$ volume densities required to collisionally excite the optically thin HCN and \hcop\ lines are about 100 times 
those required to excite the CO lines.  Hence, see Sect. 4.5, we conclude that the HCN and \hcop\ emission  does 
not trace the bulk of the gas but the denser clumps.  Combined with the derived filling factors, and dependent on 
the assumed but poorly known abundances of HCN and \hcop, we estimate (Sect. 4.5) that the gas emitting the HCN(1-0) 
and \hcop(1-0) flux represents some 10\% of the total gas mass.

5.  The critical densities of HCN and \hcop\ are not the same and we use RADEX non-LTE models to obtain estimates 
of the HCN column density and use the HCN/\hcop intensity ratio to constrain the volume density of the "dense" gas.
Because these lines are optically thin, the intensity ratio depends on the abundance ratio.
For $\chi_{\rm HCO^+} = \chi_{\rm HCN}/3$ \citep{Watanabe17}, the "dense" gas is likely $n(H_2) \la 10^4$\ccm\ 
but if the relative \hcop abundance is higher, as obtained in Section 4, then denser gas is required for the same line ratio.

6.  As part of the modeling, it was realized that the high cross-F rates initially found by \citet{benabdallah2012} could not fit our data
but that the revised rates by \citet{hernandez2017} provided a good fit.  It was thus possible to observationally test hyperfine collision rates.

7. The original motivation was to explore the link between star formation and dense gas in the poorly known outer disk environment.
Our data fall on the general $L_{HCN} - L_{IR}$ plot (Fig. \ref{f:sfr-gas2}) but, within our sample, there is no correlation 
between the SFR, as traced by the 8 and/or 24$\mu$m flux, and the 
flux in the HCN and \hcop\ lines or the dense gas fraction as traced by the HCN/CO or \hcop/CO line ratios.  
Instead, we find that the dense gas fraction decreases away from the galactic plane, presumably due to lower pressure.
An analysis of the \cite{Yuan2016} sources reveals the same trend.  

\begin{acknowledgements}
We would like to dedicate this work to our dear friend and colleague Prof. Yu Gao, who passed away during the revision of the article.
We would like to thank the DeLingHa telescope staff and the Purple Mountain Observatory for making these observations possible.  
JB would like to thank NanJing University for hosting.  We would also like to ackowledge grants from the french PNGC and the ANR program for financing through the ANR-11-BS56-010 project STARFICH.  
Y.S. acknowledges support by the Youth Innovation Promotion Association, CAS (Y2022085), {\it Light of West China} Program, CAS (No. xbzg-zdsys-202212), and the NSFC through grant 11773077.
HC is supported by Key Research Project of Zhejiang Lab?No. 2021PE0AC0.
Y.G. was supported by the National Key Basic Research and Development
Program of China (2017YFA0402700), the National Natural Science Foundation of China (11861131007, U1731237, 12033004), and Chinese Academy of
Sciences Key Research Program of Frontier Sciences (QYZDJSSW-SLH008).
This publication makes use of data products from the Wide-field Infrared Survey Explorer, which is a joint project of the University of California, Los Angeles, and the Jet Propulsion Laboratory/California Institute of Technology, funded by the National Aeronautics and Space Administration.

\end{acknowledgements}

\bibliographystyle{aa}
\bibliography{jb}

\end{document}